\documentclass{article}
\usepackage[draft]{graphicx} 
\usepackage{amsmath}
\newcommand{\D}[2]{\frac{d #1}{d #2}}

\newcommand{\iu}{{i\mkern1mu}}
\usepackage{xcolor}
\usepackage{amssymb}
\usepackage{caption}
\usepackage{placeins}
\usepackage{subcaption}
\usepackage{authblk}
\usepackage[sort,numbers]{natbib}

\title{Weak-DMD: A Galerkin approach to the problem of noise in the Dynamic Mode Decomposition algorithm  }
\author[1]{William Bennett}
\author[2]{Ryan G. McClarren}
\author[3]{Ethan Smith}
\author[4]{Melek Derman}
\affil[1]{Los Alamos National Laboratory, XCP-8: Verification and Analysis}
\affil[2]{University of Notre Dame, Department of Aerospace and Mechanical Engineering}
\affil[3]{Los Alamos National Laboratory, XTD-IDA: Integrated Design and Assessment}
\affil[4]{Oregon State University, Department of Nuclear Engineering}
\date{\today}

\begin{document}
\maketitle
\begin{abstract}
    Dynamic Mode Decomposition (DMD) is a data-driven method for approximating the spatiotemporal modes of a system. The eigenvectors and eigenvalues of the system are approximated from a series of time-snapshots of the state variables. The standard formulation of DMD is subject to strict assumptions concerning the time-spacing of the snapshots and is biased by measurement noise. Variations on the method have been developed to address these shortcomings, but the problem is still open. Motivated by the effectiveness of Galerkin methods in the field of model discovery, a weak formulation of DMD is presented, weak-DMD. Weak-DMD precludes timestep considerations and also filters noise. Results for two nuclear engineering applications and the flow of fluid past a cylinder are given and compared with a state of the art DMD algorithm.
\end{abstract}

\section{Introduction}\label{sec:intro}

Dynamic Mode Decomposition (DMD) is a popular data-driven technique for analyzing the time behavior of complex experimental and simulated systems. This method has been used to model the evolution of turbulent flows \cite{cassinelli2017streak,seena2011dynamic, schmid2010dynamic}, to calculate $\alpha$ eigenvalues for neutron transport \cite{mcclarren2019calculating, smith2023variable}, and to tame high dimensional control problems \cite{proctor2016dynamic}. DMD forms an approximation of the Koopman operator with a certain number of time snapshots of a system with the useful property that the eigenvalues and eigenvectors for this operator are relatable to the eigenvalues and eigenvectors of the original operator \cite{rowley2009spectral, mezic2013analysis}. 

DMD algorithms are traditionally restricted by the requirement that the data snapshots be equally spaced in time. $\theta$-DMD \cite{li2022dynamic} was developed to address this issue. For simulation data generated with a known time-integration method, Variable-DMD (VDMD) has shown promise \cite{smith2023variable}. DMD methods have been shown to be biased by measurement noise \cite{dawson2016characterizing}. Methods have been developed to ameliorate the effects of noise corruption, such as total-DMD (tDMD) \cite{dawson2016characterizing, hemati2017biasing}, subspace-DMD (sDMD) \cite{takeishi2017subspace}, and optimized-DMD \cite{askham2018variable}. tDMD solves a least-squares problem to de-bias the spectra. This approach accounts for bias error, not random error and also requires the dynamics of the underlying system to be deterministic. sDMD does not invoke the deterministic assumption but makes strong assumptions on the structure of the noise. Optimized-DMD poses a nonlinear least squares problem by treating all of the snapshots at once. This method requires no restrictions on the spacing of the snapshots and is noise resilient. Optimized-DMD is not guaranteed to converge, however \cite{askham2018variable}. Because the method is widely recommended \cite{brunton2021modern, tartakovsky2020physics,brunton2025machine, bagheri2013koopman} optimized-DMD will serve as the benchmark against which the performance of the method presented here will be compared in the results section.  

Taking a different track than the methods previously mentioned, we address the issues of time-spacing and noise by presenting a weak formulation of the DMD problem similar to weak-SINDY \cite{messenger2021weak} in the realm of model discovery. This method is called weak-DMD. No additional assumptions are made concerning the underlying dynamics of the system and no assumptions are made about the structure of the measurement noise. Also, weak-DMD is agnostic to the spacing of the snapshots. The dimension of the system is determined by the rank of the trial and test basis spaces, not the number of time-snapshots. While weak-DMD is equally applicable to experimental data and simulation results, the primary focus in this work will be numerical simulations.

The first problem considered is a neutronics criticality problem for which an analytic solution exists. Gaussian noise is added to the measurements artificially. Because the eigenvalues are known and the noise is well behaved, this problem serves as a suitable proof of concept. Next, Monte Carlo data for neutron transport in a more complex system is analyzed. The noise in this case is not synthetic, but due to the randomness inherent in the numerical method. This Monte Carlo problem provides a more realistic forecast of how the method will perform on non-synthetic noisy data. Lastly, to apply the method to data that is not noisy in a random sense but instead characterized by deterministic stochasticity from turbulence, data from a hydrodynamical calculation of flow past a two dimensional cylinder is treated. For each problem, results are compared with those produced by an implementation of optimized-DMD from \texttt{PyDMD} \cite{ichinaga2024pydmd,demo2018pydmd}. This implementation includes a capability for statistical averaging of an ensemble of models \cite{sashidhar2022bagging}, called ``bagging'', which is intended to promote robustness to noise.



This paper is organized as follows. In Section \ref{sec:methods} the weak formulation of DMD is derived. A simplified example is worked through in Section \ref{sec:worked} to demonstrate the method. Section \ref{sec:imp} contains practical information about the choices made in implementing the method and selecting parameters. Section \ref{sec:results} gives results for two neutronics eigenvalue problems and a flow past a cylinder problem followed by conclusions and future work in Section \ref{sec:conc}.




\section{Method Derivation}\label{sec:methods}
Suppose measurement data subject to noise of states $\boldsymbol{y}$ are given over $N$ unequally spaced time steps,
\begin{equation}
    \boldsymbol{x}_n = \boldsymbol{y}_n(t_n) + \boldsymbol{\epsilon}_n
\end{equation}
where each $\boldsymbol{x}_n$ is a vector of length $M$ and is composed of the state variables $\boldsymbol{y}$ and measurement noise $\boldsymbol{\epsilon}$. Assume that the state variables are governed by the generic linear system of ordinary differential equations (ODEs) of dimension $M$
\begin{equation}\label{eq:ode}
   \D{}{t} \boldsymbol{y}(t) = \boldsymbol{A} \boldsymbol{y}(t), \quad 0\leq t \leq T,
\end{equation}
with initial conditions 
\begin{equation}\label{eq:IC}
    \boldsymbol{y}(0) = \boldsymbol{y}_0 \in \mathbb{R}^M. 
\end{equation}
The operator $\boldsymbol{A}$ is of dimension $M\times M$. Approximation of the time eigenvalues and spatial modes of the operator $\boldsymbol{A}$ from the measurement data is the basic problem that DMD methods are intended to solve. 


To pose the system Eq.~\eqref{eq:ode} in its weak form, multiply Eq.~\eqref{eq:ode} by some time dependent test basis function with compact support, $\phi_i$,
\begin{equation}\label{eq:ode_phi}
   \phi_i(t)\D{}{t} \boldsymbol{y}(t) = \phi_i(t)\boldsymbol{A} \boldsymbol{y}(t).
\end{equation}
The number of test basis functions is $I$. Next, project the measurement data onto a trial space, $\psi_i$
\begin{equation}\label{eq:y_projection}
        \boldsymbol{x}(t)= \sum_{j=1}^J\:\boldsymbol{c}_{j} \:\psi_j(t).
\end{equation}
Where $\boldsymbol{c}$ is a matrix of rank $J\times M$ containing the expansion coefficients. Multiply both sides of Eq.~\eqref{eq:y_projection} by a trial basis and integrate, giving a matrix equation for each element of $\boldsymbol{x}$ 
\begin{equation}\label{eq:expansion_system}
    \boldsymbol{a}_m = \boldsymbol{G}\boldsymbol{c}_m
\end{equation}
where $\boldsymbol{G}$, the Gram matrix, is defined
\begin{equation}\label{eq:Gram_mat}
    G_{i, j} = \int_0^T\!dt'\:\psi_i(t')\psi_j(t').
\end{equation}
The column vector $a_m$ $[a_{1,m}, a_{2,m}, \dots a_{I,m}]^\mathrm{T}$ contains projections of each element of the data vector onto the trial basis
\begin{equation}
    a_{i,m} = \int_0^T\!dt'\:\psi_i(t')x_m(t').
\end{equation}
It has proven beneficial to solve Eq.~\eqref{eq:expansion_system} with a least squares approach as the system becomes increasingly ill conditioned as the trial basis, which is non-orthogonal in this case, grows in size.


The next step is to assume that the projection step returns a substantially de-noised representation of the measurement data, that is, $\sum_{j=1}^J\:\boldsymbol{c}_j \:\psi_j(t) \approx \boldsymbol{y}$. Evidence supporting the validity of this assumption is visible in the results of Section \ref{subsec:12grp} in Figures \ref{fig:12grp_subcrit_recon} and \ref{fig:12grp_supercrit_recon}. Substituting this expansion for $\boldsymbol{y}$ in Eq.~\eqref{eq:ode_phi} and taking the inner product gives
\begin{equation}\label{eq:intermediate_step_1}
    \left<\phi_i, \D{}{t} \sum_{j=1}^J\:\boldsymbol{c}_j \:\psi_j\right> = \left<\phi_i, \boldsymbol{A} \sum_{j=1}^J\:\boldsymbol{c}_j \:\psi_j\right>.
\end{equation}
Integration by parts of Eq.~\eqref{eq:intermediate_step_1} yields, 
\begin{equation}\label{eq:final_step}
    \phi_i \:\sum_{j=1}^J\: \boldsymbol{c}_j\psi_j\bigg{|}_{t_1}^{t_2}-\left<\D{\phi_i}{t}, \sum_{j=1}^J\:\boldsymbol{c}_j \:\psi_j\right> =
    \boldsymbol{A} \sum_{j=1}^J\boldsymbol{c}_j \left<\phi_i,  \:\psi_j\right>.
\end{equation}
Unlike the inner product that is calculated to determine the basis expansion coefficients, the inner product in Eq.~\eqref{eq:final_step} is over the selected time window, $[t_1, t_2]$. Note that estimating $\boldsymbol{y}$ with a basis expansion has allowed the operator to be taken out of the inner product. We have not made the common requirement that the trial basis be zero at the boundary. Trial bases are allowed to span the boundaries of the time domain to better represent the dynamics of the system at those extrema. In some cases, this results in basis functions having supports that span into negative time. This is not an issue as the expansion is never evaluated outside of the time window $t\in [t_1,t_2]$. The nonzero basis at the boundaries results in a correction term that requires evaluating the basis at the initial and final times. 

Eq.~\eqref{eq:final_step} can be written in the form,
\begin{equation}\label{eq:final_matrix_eq}
    \boldsymbol{Y}^+ = \boldsymbol{A}\boldsymbol{Y}^-,
\end{equation}
where the matrices $\boldsymbol{Y}^+$ and $\boldsymbol{Y}^-$ are matrices of dimension $M \times I$ with $I$ being the number of test bases. They are defined
\begin{equation}\label{eq:Yplus}
    \boldsymbol{Y}^+_i = \phi_i \:\sum_{j=1}^J\: \boldsymbol{c}_j\psi_j\bigg{|}_{0}^{t_f}-\sum_{j=1}^J\:\boldsymbol{c}_j\left<\D{\phi_i}{t},  \:\psi_j\right>,
\end{equation}
and
\begin{equation}\label{eq:Yminus}
    \boldsymbol{Y}^-_i = \sum_{j=1}^J\boldsymbol{c}_j \left<\phi_i,  \:\psi_j\right>.
\end{equation}
Finally, DMD may be applied to Eq.~\eqref{eq:final_matrix_eq} to estimate the eigenvalues of $\boldsymbol{A}$. First, apply a thin singular value decomposition (SVD) to $\boldsymbol{Y}^-$,
\begin{equation}\label{eq:thin_SVD}
    \boldsymbol{Y}^-=\boldsymbol{LSR}^T,
\end{equation}
where $\boldsymbol{L}$ has dimensions $L\times r$ with $r$ the number of singular values. $S$ is a diagonal matrix containing the singular values. $\boldsymbol{R}$ has dimensions $I\times r$. 

$\boldsymbol{L}$ right multiplied by its transpose is the identity matrix as is $R$ left multiplied by its transpose. Using these properties, right multiply Eq.~\eqref{eq:final_matrix_eq} by $\boldsymbol{RS}^{-1}$ and then left multiply by $\boldsymbol{L}^T$ and obtain
\begin{equation}\label{eq:Atilde}
    \boldsymbol{L}^T\boldsymbol{Y}^+\boldsymbol{RS}^{-1} = \boldsymbol{L}^T\boldsymbol{A}\boldsymbol{L}\equiv \widetilde{\boldsymbol{A}},
\end{equation}
where $\widetilde{\boldsymbol{A}}$ is an $r\times r$ lower dimensional approximation to $\boldsymbol{A}$. It can be shown that the eigenvalues and eigenvectors of $\widetilde{\boldsymbol{A}}$ are also eigenvalues and eigenvectors of $\boldsymbol{A}$ \cite{mcclarren2019calculating, tu2013dynamic}. Therefore, $\widetilde{\boldsymbol{A}}$ may be used to predict future dynamics of the system. The solution $\Delta t$ into the future is 
\begin{equation}
    \left(\boldsymbol{I}-\Delta t \widetilde{\boldsymbol{A}}_f\right)\boldsymbol{y}(t+\Delta t)  = \boldsymbol{y}(t),
\end{equation}
where $\widetilde{\boldsymbol{A}}_f$ is $\widetilde{\boldsymbol{A}}$ projected back onto the full space,
\begin{equation}\label{eq:Atilde_full}
    \widetilde{\boldsymbol{A}}_f = \boldsymbol{L}\:\widetilde{\boldsymbol{A}}\boldsymbol{\overline{L}}^T.
\end{equation}
The overline denotes the conjugate. The spatial modes are calculated 
\begin{equation}\label{eq:spatial_modes}
    \Phi = \boldsymbol{Y}^+ \boldsymbol{R}\boldsymbol{S}^{-1}\boldsymbol{W}
\end{equation}
where $\boldsymbol{W}$ are the eigenvectors of $\widetilde{\boldsymbol{A}}$. $\boldsymbol{R}$ and $\boldsymbol{S}^{-1}$ are both truncated to the range of $\widetilde{\boldsymbol{A}}$.




\section{Demonstration of a simple problem}\label{sec:worked}
For illustrative purposes, first consider the system

\begin{align}
        \D{}{t} \begin{bmatrix}
           y_{0} \\
           y_{1} \\
         \end{bmatrix}  = 
          \begin{bmatrix}
           0 \:\:\:\:\:\:\:1 \\
           -\omega^2 \:-\alpha \\
         \end{bmatrix}\begin{bmatrix}
           y_{0} \\
           y_{1} \\
         \end{bmatrix}, 
\end{align}  
with $y_0(0) = 1, y_1(0) = 0$. The solution is 
\begin{align}
    y_0(t)&= e^{-\frac{\alpha t}{2}} \left(\frac{\alpha \sinh \left(\frac{1}{2} t \sqrt{\alpha ^2-4 \omega ^2}\right)}{\sqrt{\alpha ^2-4 \omega ^2}}+\cosh \left(\frac{1}{2} t \sqrt{\alpha ^2-4 \omega ^2}\right)\right),\\y_1(t)&= -\frac{2 \omega ^2 e^{-\frac{\alpha t}{2}} \sinh \left(\frac{1}{2} t \sqrt{\alpha ^2-4 \omega ^2}\right)}{\sqrt{\alpha ^2-4 \omega ^2}}. 
\end{align}
The eigenvalues of the operator are 
\begin{equation}
    \lambda = \frac{-\alpha \pm \sqrt{\alpha^2 -4 \omega^2}}{2}.
\end{equation}
For the parameters $\omega = \frac{13\sqrt{29}}{20}, \alpha = \frac{1}{10}$, the eigenvalues are
\begin{equation}\label{eqs:exact_eigs_toy}
    \lambda = -\frac{1}{20} \pm \frac{7}{2}i.
\end{equation}

Take $n$ measurements $\boldsymbol{x}$, of the system, with Gaussian noise, $\eta$ having zero mean and standard deviation $\sigma$,
\begin{equation}
    \boldsymbol{x}_n = \boldsymbol{y}(t_n) + \boldsymbol{\eta}(t_n).
\end{equation}
The first step in the weak DMD process is to project the solution onto the basis by solving Eq.~\eqref{eq:expansion_system}. Although the weak-DMD has been implemented with a finite element approach where the basis functions are compactly supported polynomials, for the example we will use a global basis and make the very fortuitous choice in selecting $y_0$ and $y_1$ as trial basis functions. This is done for the sake of illustration. 

For two trial functions, the projection coefficients are 
\begin{align}
    \boldsymbol{a}_0 = \begin{bmatrix}
        \int_{t_1}^{t_2}\!dt'\:x_0(t)\psi_0(t) \\\int_{t_1}^{t_2}\!dt'\:x_1(t)\psi_0(t)
    \end{bmatrix},
\end{align}
and
\begin{align}
    \boldsymbol{a}_1 = \begin{bmatrix}
        \int_{t_1}^{t_2}\!dt'\:x_0(t)\psi_1(t) \\\int_{t_1}^{t_2}\!dt'\:x_1(t')\psi_1(t')
    \end{bmatrix} .
\end{align}
To show the effect of the noise on the projection coefficient, write out the coefficient corresponding to the projection of the zeroth basis function onto the first equation $(y_0)$ in the system
\begin{equation}\label{eq:zerozero_coeff}
    a_{0,0} = \int_{t_1}^{t_2}\!dt'\left[\:\psi_0(t')y_0(t') + \eta(t') \psi_0(t)\right].
\end{equation}
Because the noise is zero mean, the integral $\int_{t_1}^{t_2}\!dt'\:\eta(t') \psi_0(t) =0$. For a numerical integrator to correctly approximate the noise integral as zero an infinite number of samples are required. Let the error incurred by taking a finite number of samples be $O(\delta),$ which will be assumed to be greater than the level of the error of the numerical integration routine. 

Separating the noise corrupted part gives
\begin{align}
    \boldsymbol{a}_0 = \begin{bmatrix}
\langle y_0, \psi_0\rangle +O(\delta)\\\langle y_1,\psi_0\rangle +O(\delta)
    \end{bmatrix},
\end{align}
and
\begin{align}
    \boldsymbol{a}_1 = \begin{bmatrix}
        \langle y_0, \psi_1\rangle +O(\delta)\\\langle y_1,\psi_1\rangle +O(\delta)
    \end{bmatrix},
\end{align}
where the inner product is defined 
\begin{equation}
    \langle f,g\rangle =  \int_{t_1}^{t_2}\!dt'\:f(t')g(t').
\end{equation}

Populating the Gram matrix $\boldsymbol{G}$ with Eq.~\eqref{eq:Gram_mat} 
\begin{align}
    \boldsymbol{G}  = \begin{bmatrix}
         \langle \psi_0, \psi_0\rangle\:\:  \langle \psi_0, \psi_1\rangle \\  \langle \psi_1, \psi_0\rangle\:\: \langle \psi_1, \psi_1\rangle
    \end{bmatrix}.
\end{align}
Solving for the expansion coefficients for the first equation with Eq.~\eqref{eq:expansion_system}
\begin{align}\label{eq:matrix_eq_c0}
     \begin{bmatrix}
         \langle \psi_0, \psi_0\rangle\:\:  \langle \psi_0, \psi_1\rangle \\  \langle \psi_1, \psi_0\rangle\:\: \langle \psi_1, \psi_1\rangle\end{bmatrix} \boldsymbol{c}_0 = \begin{bmatrix}
\langle y_0, \psi_0\rangle +O(\delta)\\\langle y_1,\psi_0\rangle +O(\delta)
    \end{bmatrix}.
\end{align}
Separating the noise component from the projection, $\boldsymbol{c}_0$ is split into exact and noise corrupted parts, $\boldsymbol{c}^{(1)}_0 + \boldsymbol{c}^{(2)}_0$, and Eq.~\eqref{eq:matrix_eq_c0} can be written
\begin{align}\label{eq:matrix_eq_c0_1}
     \begin{bmatrix}
         \langle \psi_0, \psi_0\rangle\:\:  \langle \psi_0, \psi_0\rangle \\  \langle \psi_1, \psi_0\rangle\:\: \langle \psi_1, \psi_1\rangle\end{bmatrix} \boldsymbol{c}_0^{(1)} = \begin{bmatrix}
\langle y_0, \psi_0\rangle \\\langle y_1,\psi_0\rangle 
    \end{bmatrix},
\end{align}
\begin{align}\label{eq:matrix_eq_c0_2}
     \begin{bmatrix}
         \langle \psi_0, \psi_0\rangle\:\:  \langle \psi_0, \psi_1\rangle \\  \langle \psi_1, \psi_0\rangle\:\: \langle \psi_1, \psi_1\rangle\end{bmatrix} \boldsymbol{c}_0^{(2)} = \begin{bmatrix}
O(\delta)\\O(\delta)
    \end{bmatrix}.
\end{align}
Now, choose the test set to equal the trial set, $\psi_0=y_0, \psi_1=y_1$. In this case, it is clear from Eqs.~\eqref{eq:matrix_eq_c0_1} and \eqref{eq:matrix_eq_c0_2} that the solution is 
\begin{align}
    \boldsymbol{c}_0 = \begin{bmatrix}
        1\: 0
    \end{bmatrix}^\mathrm{T} + \begin{bmatrix}
O(\delta)\:O(\delta)
    \end{bmatrix}^\mathrm{T}.
\end{align}
Similarly, the expansion coefficients for the second equation are
\begin{align}
    \boldsymbol{c}_1 = \begin{bmatrix}
        0\: 1
    \end{bmatrix}^\mathrm{T} + \begin{bmatrix}
O(\delta)\:O(\delta)
    \end{bmatrix}^\mathrm{T}.
\end{align}
Now it is possible to form the DMD snapshot matrices with Eqs.~\eqref{eq:Yminus} and Eq.~\eqref{eq:Yplus}. $\boldsymbol{Y}^-$ is formed by taking the inner product of the trial expansion with the test basis, $\phi$
\begin{align}
    \boldsymbol{Y}_0^- = \begin{bmatrix}
        1\: 0
    \end{bmatrix}^\mathrm{T} \langle \phi_0, \psi_0 \rangle + \begin{bmatrix}
        0\: 1
    \end{bmatrix}^\mathrm{T} \langle \phi_0, \psi_1\rangle + \begin{bmatrix}
O(\delta)\:O(\delta)
    \end{bmatrix}^\mathrm{T} ,
\end{align}
.\begin{align}
    \boldsymbol{Y}_1^- = \begin{bmatrix}
        1\: 0
    \end{bmatrix}^\mathrm{T} \langle \phi_1, \psi_0 \rangle + \begin{bmatrix}
        0\: 1
    \end{bmatrix}^\mathrm{T} \langle \phi_1, \psi_1\rangle+ \begin{bmatrix}
O(\delta)\:O(\delta)
    \end{bmatrix}^\mathrm{T}.
\end{align} 
$\boldsymbol{Y}^+$, which transfers the action of the derivative onto the test basis is defined by Eq.~\eqref{eq:Yplus}

\begin{align}
    \boldsymbol{Y}_0^+  =\phi_0(t_2)\boldsymbol{f}\bigg{|}_{t=t_2} -\phi_0(t_1)\boldsymbol{f}\bigg{|}_{t=t_1}  -\begin{bmatrix}
        1\: 0
    \end{bmatrix}^\mathrm{T} \langle \phi'_0, \psi_0 \rangle - \begin{bmatrix}
        0\: 1
    \end{bmatrix}^\mathrm{T} \langle \phi'_0 ,\psi_1\rangle + \begin{bmatrix}
O(\delta)\:O(\delta)
    \end{bmatrix}^\mathrm{T},
\end{align}
\begin{align}
    \boldsymbol{Y}_1^+  =\phi_1(t_2)\boldsymbol{f}\bigg{|}_{t=t_2} -\phi_1(t_1)\boldsymbol{f}_{t=t_1}  -\begin{bmatrix}
        1\: 0
    \end{bmatrix}^\mathrm{T} \langle \phi'_1, \psi_0 \rangle - \begin{bmatrix}
        0\: 1
    \end{bmatrix}^\mathrm{T} \langle \phi'_1 ,\psi_1\rangle + \begin{bmatrix}
O(\delta)\:O(\delta)
    \end{bmatrix}^\mathrm{T}.
\end{align}
Where $\boldsymbol{f}$ is the weak form of the solution,
\begin{equation}
    \boldsymbol{f} = \begin{bmatrix}
        1\: 0
    \end{bmatrix}^\mathrm{T} \psi_0 + \begin{bmatrix}
        0\: 1
    \end{bmatrix}^\mathrm{T} \psi_1. 
\end{equation}
Next, choose the test basis to be identical to the trial. In this case, the matrices $\boldsymbol{Y}^+$ and $\boldsymbol{Y}^-$ simplify to
\begin{align}\label{eq:Yminus-toy-final}
    \boldsymbol{Y}^- = \begin{bmatrix}
        \langle y_0, y_0\rangle \: \langle y_0, y_1\rangle \\\langle y_1, y_0\rangle \: \langle y_1, y_1\rangle
    \end{bmatrix} + \boldsymbol{O(\delta)},
\end{align}
\begin{align}\label{eq:Yplus-toy-final}
    \boldsymbol{Y}^+ = \begin{bmatrix}
        y_0y_0\bigg{|}_{t_1}^{t_2} - \langle y_0',y_0\rangle \:\:\:\: y_0y_1\bigg{|}_{t_1}^{t_2} - \langle y_0',y_1\rangle \\ 
        y_1,y_0\bigg{|}_{t_1}^{t_2} - \langle y_1',y_0\rangle \:\:\:\: y_1y_1\bigg{|}_{t_1}^{t_2} - \langle y_1',y_1\rangle 
    \end{bmatrix}+\boldsymbol{O(\delta)}.
\end{align}
$\widetilde{\boldsymbol{A}}$ is formed by right multiplying $\boldsymbol{Y}^+$ by $(\boldsymbol{Y}^-)^{-1}$. While this operation is tractable for the system we consider, and even the inner products have analytic solutions, it is not necessary for our purposes to detail this step. It is important to point out, however, that the noise term carries in an additive way, that is
\begin{equation}
    \widetilde{\boldsymbol{A}} = \boldsymbol{Y}^+(\boldsymbol{Y}^-)^{-1} + \boldsymbol{O(\delta)}. 
\end{equation}
This result suggests that the amount of noise corruption that survives the trial projection step corrupts the approximated operator $\widetilde{\boldsymbol{A}}$. As mentioned earlier, the weak projection step is meant to reduce the influence of noise and it is assumed that $O(\delta) < O(\eta)$.

For zero noise, a table of eigenvalues found by evaluating the inner products in Eqs.~\eqref{eq:Yminus-toy-final} and \eqref{eq:Yplus-toy-final}, forming $\widetilde{\boldsymbol{A}}$, and solving the eigenvalue problem is given in Table \ref{tab:toy_prob_vals}. These results show that even for optimal test and trial bases, a sufficiently large integration domain window is required to learn the dynamics of the system. 

\begin{table}[]
\begin{tabular}{l|ll}
$t_2$ & $\lambda_1 = -0.05 + 3.5 \iu$            & $\lambda_2 = -0.05 -3.5\iu$            \\ \hline
1     & $-0.483125+36.9641 \iu$  & $-0.483125-36.9641 \iu$  \\
5     & $-0.128588+8.90838\iu$   & $-0.128588-8.90838\iu$   \\
20    & $-0.0577018+4.04781 \iu$ & $-0.0577018-4.04781 \iu$ \\
100   & $-0.0500021+3.5\iu$      & $-0.0500021-3.5 \iu$    
\end{tabular}
\caption{Eigenvalues calculated for the toy example of Section \ref{sec:worked} setting $t_1=0$ and varying $t_2$ in Eqs. ~\eqref{eq:Yminus-toy-final} and \eqref{eq:Yplus-toy-final}. To be compared with the exact values given by Eq.~\eqref{eqs:exact_eigs_toy}.}
\label{tab:toy_prob_vals}

\end{table}

\section{Implementation}\label{sec:imp}
The test and trial basis are of the form used by \cite{messenger2021weak},
\begin{equation}
    \phi(t) = \begin{cases}
        C(t-a)^p(b-t)^p & t\in (a,b)\\
        0& \mathrm{otherwise}
    \end{cases} 
\end{equation}
where the normalization constant is defined
\begin{equation}
    C = \left(\frac{2}{b-a}\right)^{2p}.
\end{equation}
The exponent, $p\geq 1$ and $a<b$. 

Choosing the optimal supports, polynomial orders, and number of the basis functions in the test and trial spaces is an open ended problem. There are many methodologies for informing these choices in the literature. Trial bases are usually selected by minimizing some empirical risk metric, as in \cite{craven1978smoothing, donoho1995adapting}. In \cite{messenger2021weak}, the center of the basis support is selected by sampling from a cumulative distribution function of the data velocities, the degree by an input parameter, and the supports by requiring that the basis is zero at the endpoints.

In our current implementation, the trial basis is constructed in a non-automated fashion and parameters are hand tuned to reduce the reconstruction noise. First, the user inputs a list of numbers and overlap percentages. For example, if an element in the number of basis lists is $2$ and the corresponding overlap percentage is $0.5$, two basis functions are added to the set with an overlap width equal to half the distance between the center of two functions. The centers of the support are a uniform grid on $[t_1,t_2]$. It has proven beneficial to allow basis supports to extend beyond the time window, even into negative time. The polynomial order $p$ is also a user specified variable. In future iterations of this work, priority will be given to automating the trial basis construction. The current method is sufficient for the present purpose however. 

For the construction of the test basis, first a basis is formed similarly to the trial. In general, the test set is larger than the trial set. Increasing the trial set without bound has the effect of interpolating the noise. This is not a concern with the test set. The matrices $\boldsymbol{Y}^+$ and $\boldsymbol{Y}^-$ are formed according to Eqs.~\eqref{eq:Yplus} and \eqref{eq:Yminus}, respectively. A SVD is performed Eq.~\eqref{eq:thin_SVD} with the number of retained eigenvalues in $\boldsymbol{S}$ determined by an energy criteria. For some value  $\mathcal{E}\leq 1$, the rank $r$ of the SVD is determined by finding the smallest $r$ such that
\begin{equation}\label{eq:DMD_energy}
    \sum_{i=0}^r\: \frac{s_i}{\overline{s}} \geq \mathcal{E} 
\end{equation}
where $s_i$ are the singular values sorted from largest to smallest and $\overline{s}$ is the total sum of the singular values. This method of choosing the SVD rank is known as the \textit{energy method.}

\section{Results}\label{sec:results}
\subsection{Twelve group criticality problem}\label{subsec:12grp}
We apply the method to synthetic data generated from a prompt neutron problem. Neutron populations in a sphere of $99$ at.$\%$ $^{239}\mathrm{Pu}$ and $1$ at.$\%$ natural carbon are approximated with a $12$ group energy discretization, meaning the energy of the neutrons will fall into one of $12$ bins, and a simple buckling model for leakage. By changing the radius of the sphere, subcritical or supercritical solutions are produced. A supercritical system requires an eigenvalue $>0$, while a subcritical system only has decay modes.

\begin{figure}
    \centering
    \includegraphics[draft = false, width=0.5\linewidth]{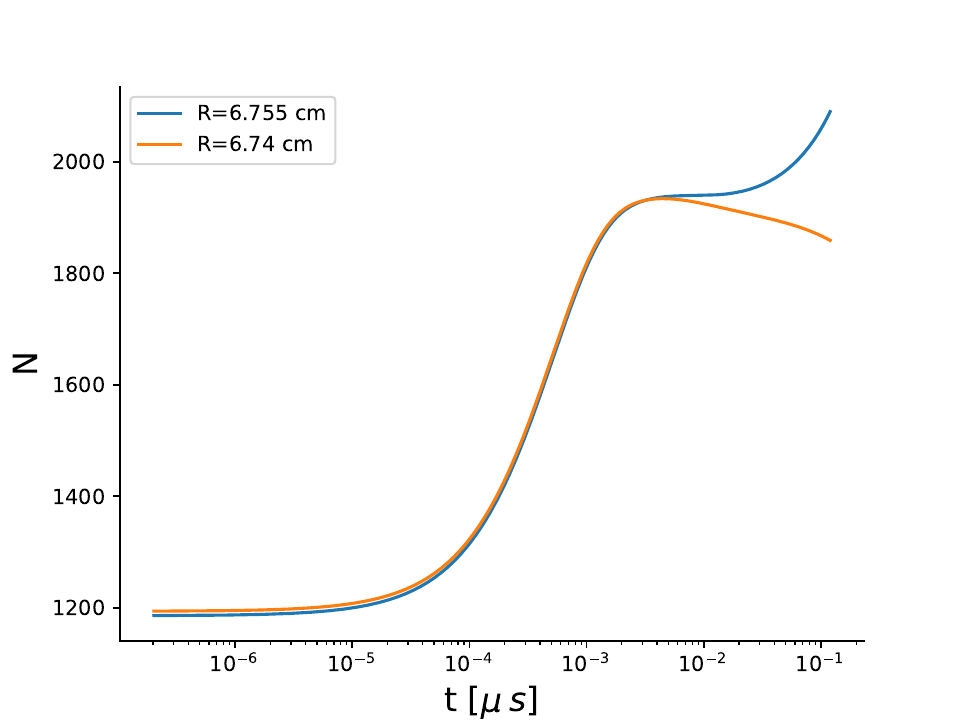}
    \caption{Group summed neutron populations versus time for two radii for the $12$ group sphere.}
    \label{fig:12grp_ode_sols}
\end{figure}

For this problem, one subcritical and one supercritical radii are selected. The transport operator is formed explicitly, therefore the eigenvalues are known analytically. The first two eigenvalues for the radii selected are given in table \ref{tab:12grp_params}. Neutron population solutions summed over all groups are shown in Figure \ref{fig:12grp_ode_sols}.

\begin{table}[]
\begin{tabular}{l|ll}
             & $\lambda_1$ & $\lambda_2$ \\ \hline
$R=6.74$ cm  & $-0.002244$ & $-0.27054$ \\
$R=6.755$ cm & $0.007565$  & $-0.270383$
\end{tabular}
\caption{The first two eigenvalues for the twelve group sphere for a subcritical and a supercritical configuration. }
\label{tab:12grp_params}
\end{table}

\begin{figure}[h!]
    \centering
    \begin{subfigure}[b]{0.49\textwidth}
        \includegraphics[draft=false,width=\textwidth]{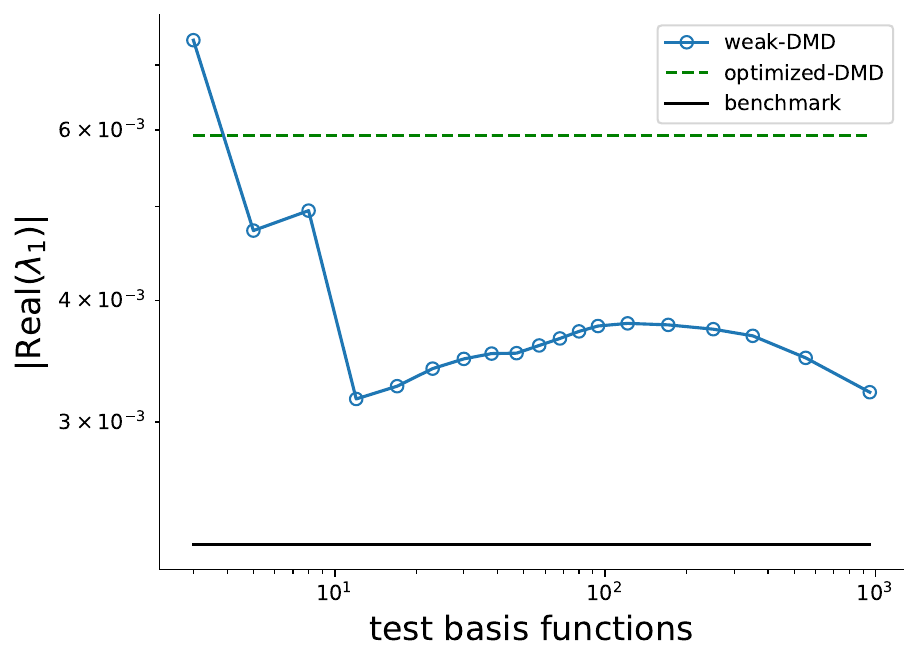}
        \caption{$\lambda_1$}
    \end{subfigure}
    \hfill 
    \begin{subfigure}[b]{0.49\textwidth}
        \includegraphics[draft=false,width=\textwidth]{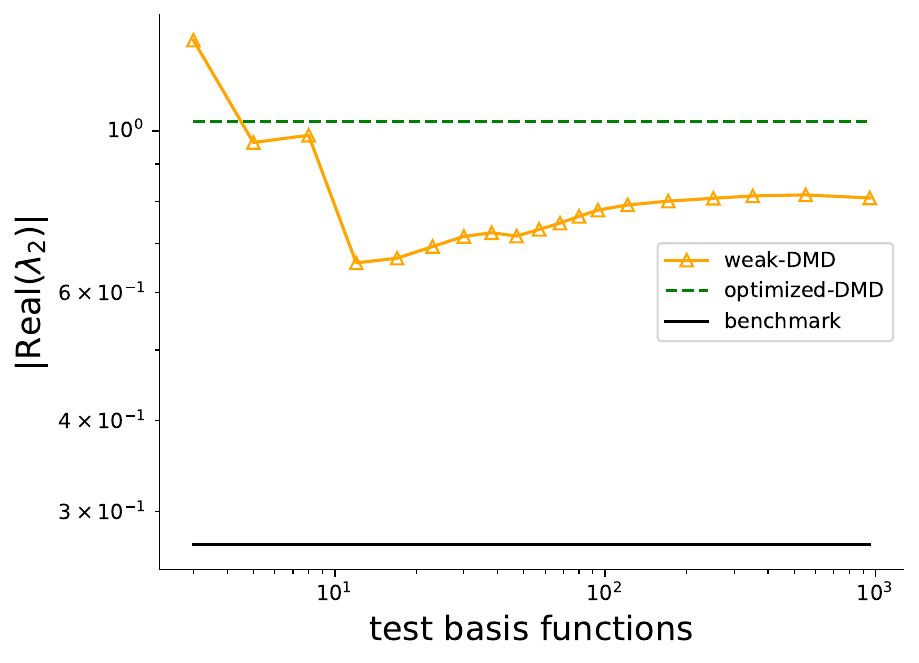}
        \caption{$\lambda_2$}
    \end{subfigure}
    \caption{Logscaled convergence of the first and second eigenvalues given by weak-DMD for the $12$ group sphere for $R=7.74\:[\mathrm{cm}]$. The benchmark solution is shown by the black line and the estimate from optimized-DMD retaining $4$ singular values and with $10$ trials in the bagging routine, by the green. The energy used in Eq.~\eqref{eq:DMD_energy} was $0.99997$ and the polynomial order $p=2$. $43$ bases were used in the trial projection. }
    \label{fig:subcrit_12_convergence_nonoise}
\end{figure}

\begin{figure}[h!]
    \centering
    \begin{subfigure}[b]{0.49\textwidth}
        \includegraphics[draft = false,width=\textwidth]{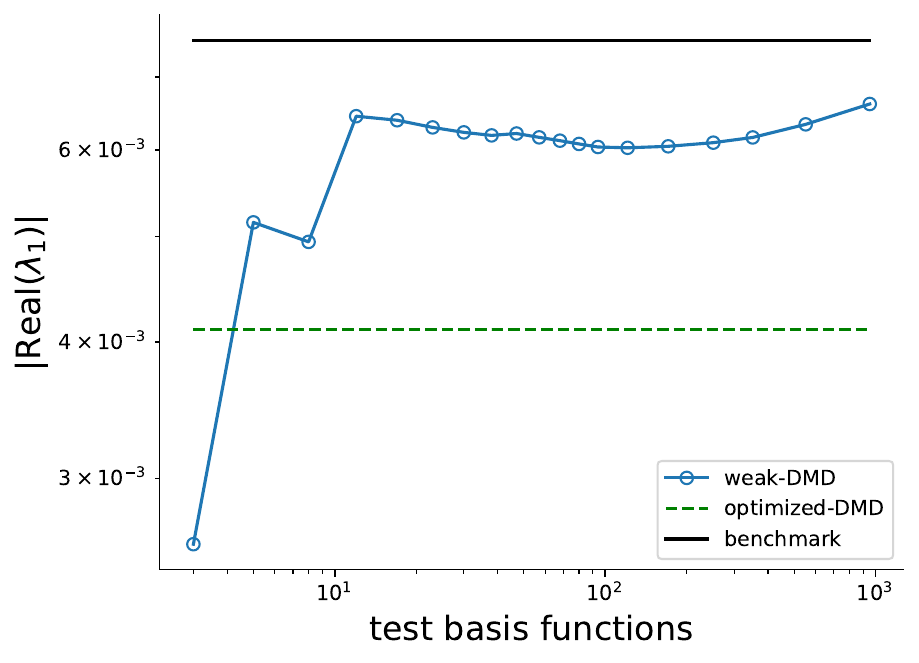}
        \caption{$\lambda_1$}
    \end{subfigure}
    \hfill 
    \begin{subfigure}[b]{0.49\textwidth}
        \includegraphics[draft=false,width=\textwidth]{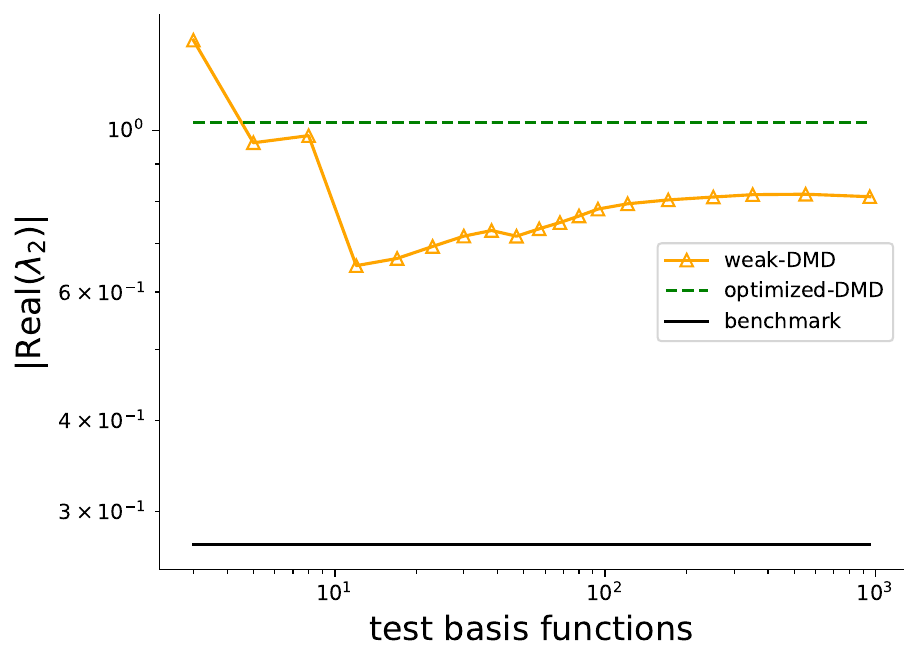}
        \caption{$\lambda_2$}
    \end{subfigure}
    \caption{Logscaled convergence of the first and second eigenvalues given by weak-DMD for the $12$ group sphere for $R=7.755\:[\mathrm{cm}]$. The benchmark solution is shown by the black line and the estimate from optimized-DMD retaining $4$ singular values and with $10$ trials in the bagging routine, by the green. The energy used in Eq.~\eqref{eq:DMD_energy} was $0.99997$ and the polynomial order $p=2$. $43$ bases were used in the trial projection. }
    \label{fig:supercrit_12_convergence_nonoise}
\end{figure}

Over the time window $t\in [0.0022, 0.009]\:[\mu\mathrm{s}]$, the eigenvalues obtained with weak-DMD are compared with optimized-DMD results in Figures \ref{fig:subcrit_12_convergence_nonoise} and \ref{fig:supercrit_12_convergence_nonoise} for the subcritical and supercritical cases, respectively. The number of trial functions remained fixed while the test basis increased. The optimized-DMD results were calculated with a SVD rank of $4$ and $10$ trials in the bagging routine. These results show the weak-DMD algorithm tending towards converged eigenvalues as the number of test functions increases for a given energy tolerance in Eq.~\eqref{eq:DMD_energy}.

\begin{figure}
    \centering
    \includegraphics[draft=false,width=0.75\linewidth]{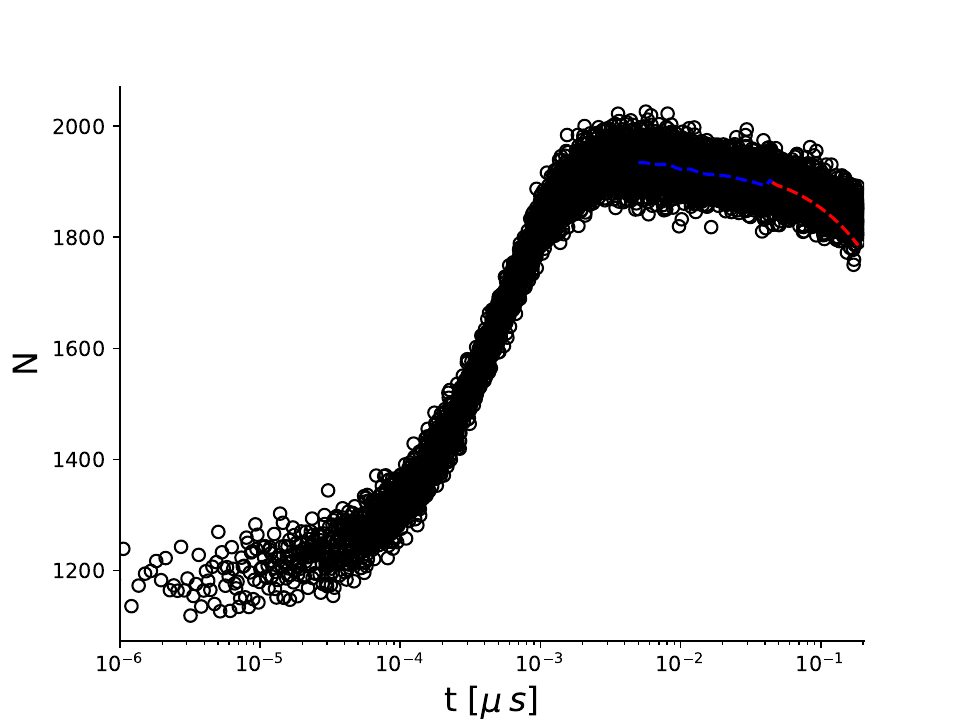}
    \caption{Measurement data (black circles) of the summed neutron population for the $12$ group sphere with $R=6.74\: [\mathrm{cm}]$. The weak-DMD solution reconstruction is shown in blue and the extrapolation in red. $14$ bases of order $p=2$ were used in the trial projection, and the test space corresponded to the last entry in Figure \ref{fig:subcrit_12_convergence_noise}. The energy used in Eq.~\eqref{eq:DMD_energy} was $0.99941$.   }
    \label{fig:12grp_subcrit_recon}
\end{figure}

\begin{figure}
    \centering
    \includegraphics[draft=false,width=0.75\linewidth]{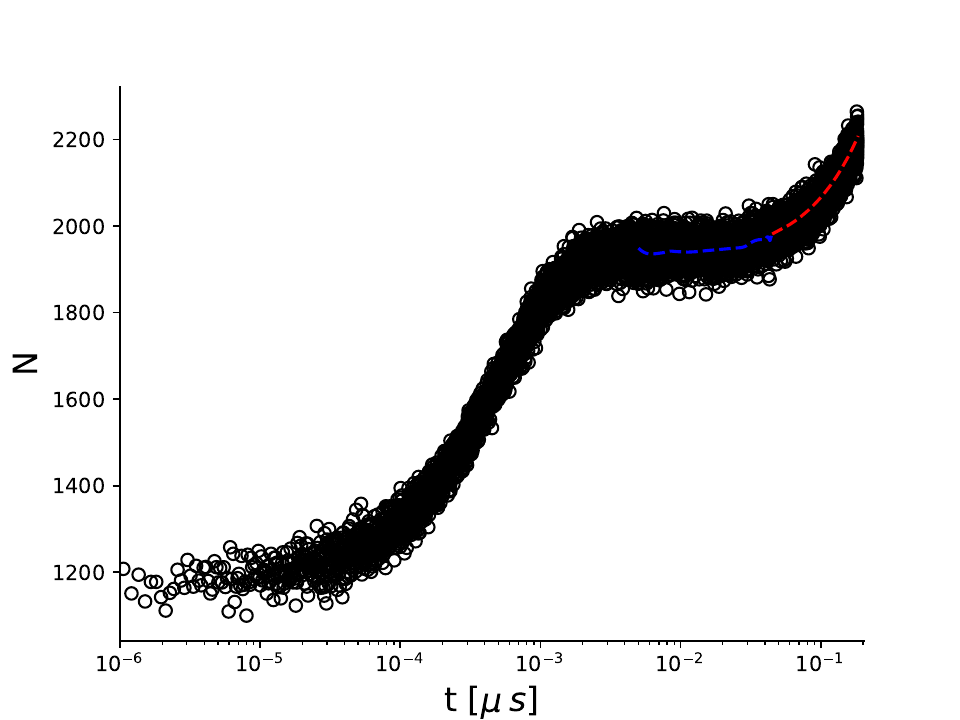}
    \caption{Measurement data (black circles) of the summed neutron population for the $12$ group sphere with $R=6.755\: [\mathrm{cm}]$. The weak-DMD solution reconstruction is shown in blue and the extrapolation in red. $14$ bases of order $p=2$ were used in the trial projection, and the test space corresponded to the last entry in Figure \ref{fig:subcrit_12_convergence_noise}. The energy used in Eq.~\eqref{eq:DMD_energy} was $0.99967$.   }
    \label{fig:12grp_supercrit_recon}
\end{figure}

\begin{figure}[h!]
    \centering
    \begin{subfigure}[b]{0.49\textwidth}
        \includegraphics[draft=false,width=\textwidth]{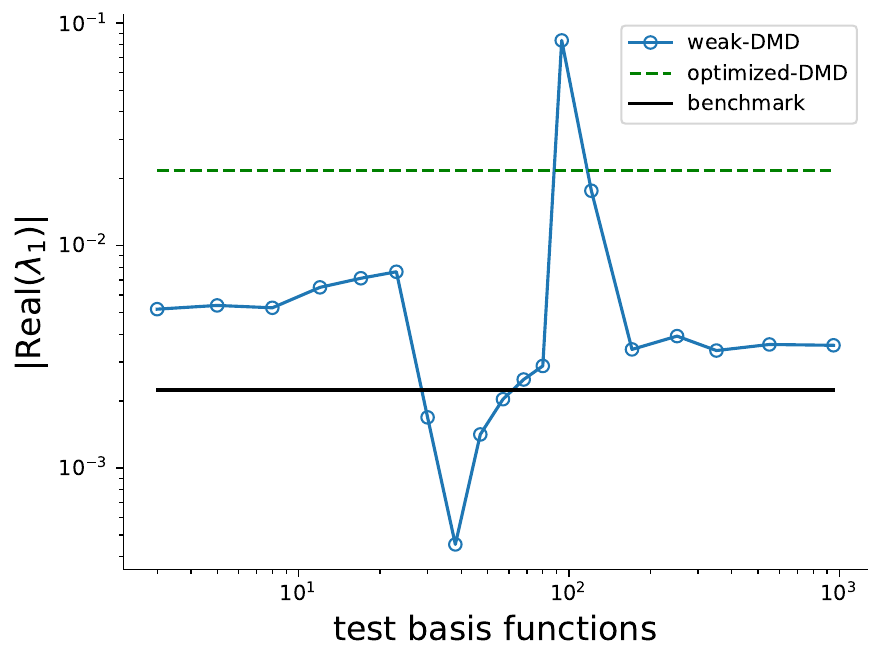}
        \caption{$\lambda_1$}
    \end{subfigure}
    \hfill 
    \begin{subfigure}[b]{0.49\textwidth}
        \includegraphics[draft=false,width=\textwidth]{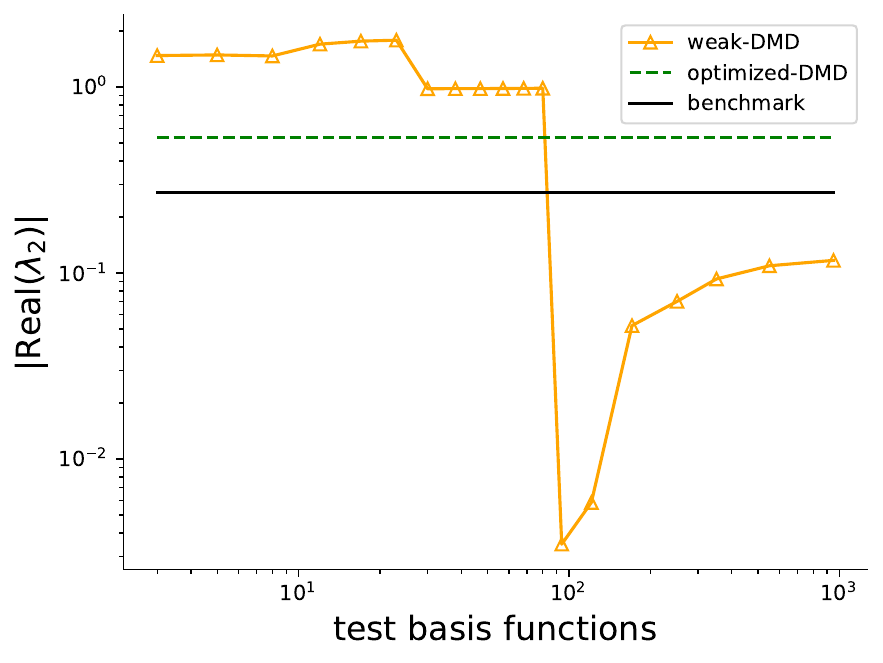}
        \caption{$\lambda_2$}
    \end{subfigure}
    \caption{Logscaled convergence of the first and second eigenvalues given by weak-DMD for the $12$ group sphere for $R=6.74\:[\mathrm{cm}]$ with synthetic noise added to the measurements. The benchmark solution is shown by the black line and the estimate from optimized-DMD retaining $4$ singular values and with $10$ trials in the bagging routine, by the green. The energy used in Eq.~\eqref{eq:DMD_energy} was $0.99945$ and the polynomial order $p=2$. $14$ bases were used in the trial projection. }
    \label{fig:subcrit_12_convergence_noise}
\end{figure}

\begin{figure}[h!]
    \centering
    \begin{subfigure}[b]{0.49\textwidth}
        \includegraphics[draft=false,width=\textwidth]{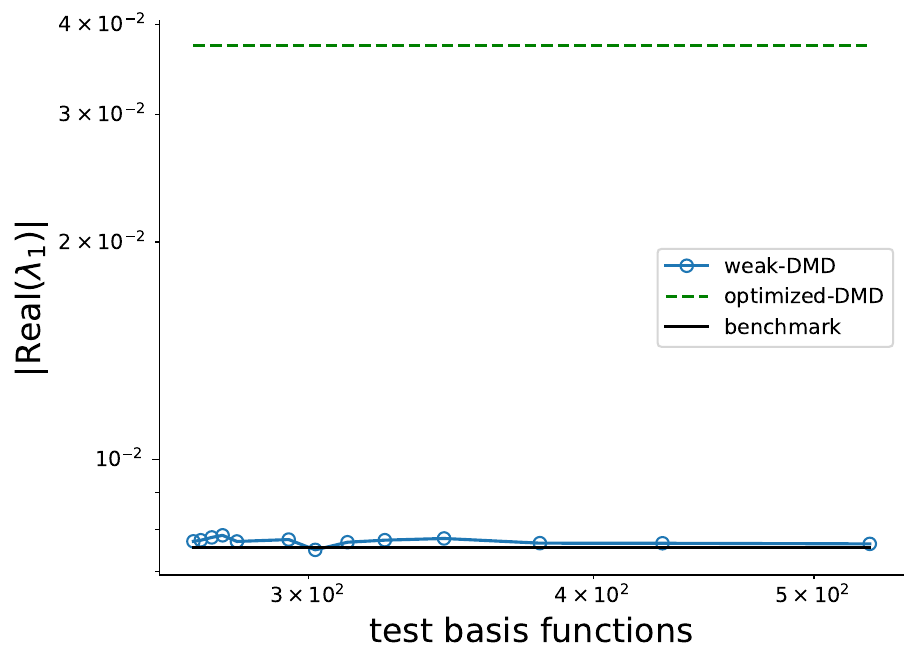}
        \caption{$\lambda_1$}
    \end{subfigure}
    \hfill 
    \begin{subfigure}[b]{0.49\textwidth}
        \includegraphics[draft=false,width=\textwidth]{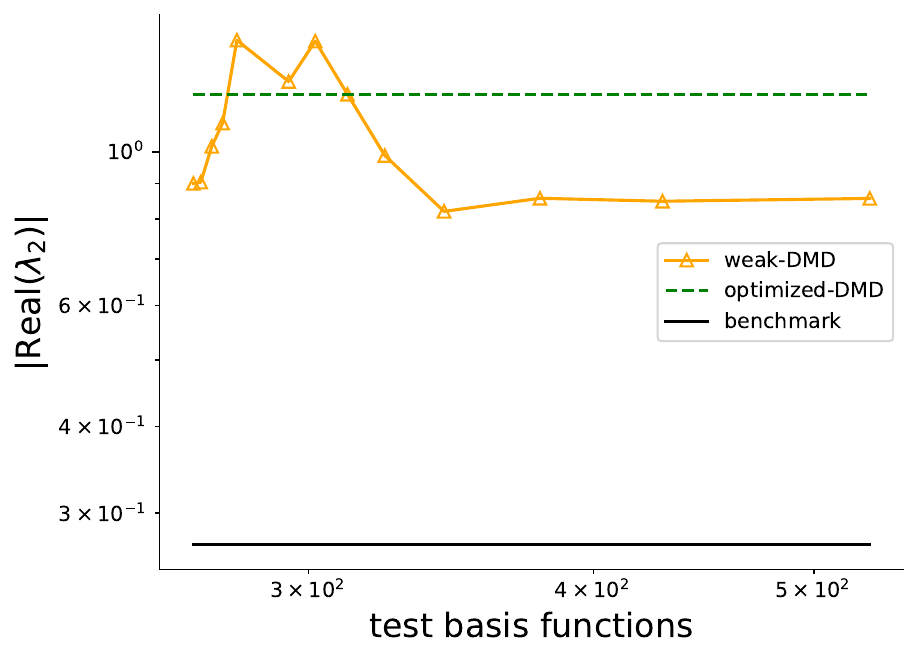}
        \caption{$\lambda_2$}
    \end{subfigure}
    \caption{Logscaled convergence of the first and second eigenvalues given by weak-DMD for the $12$ group sphere for $R=6.755\:[\mathrm{cm}]$ with synthetic noise added to the measurements. The benchmark solution is shown by the black line and the estimate from optimized-DMD retaining $4$ singular values and with $10$ trials in the bagging routine, by the green. The energy used in Eq.~\eqref{eq:DMD_energy} was $0.99967$ and the polynomial order $p=2$. $14$ bases were used in the trial projection. }
    \label{fig:supercrit_12_convergence_noise}
\end{figure}

Synthetic Gaussian noise was then added to the data with a variance of $\sigma = 0.2$ and a magnitude that is $15\%$ of the neutron population at a given time. Measurement data and the reconstructions and future projections from weak-DMD, are plotted in Figures \ref{fig:12grp_subcrit_recon} and \ref{fig:12grp_supercrit_recon}. Convergence plots for the noised data as the test space is augmented are shown in Figures \ref{fig:subcrit_12_convergence_noise} and \ref{fig:supercrit_12_convergence_noise}. 

For these results, the convergence of the eigenvalues as the test space grows seems slightly more erratic than the noiseless results. It was necessary to reduce the size of the trial basis from $43$ to $14$ to avoid overfitting and also to decrease the energy of the retained singular values.


While the results show weak-DMD outperforming optimized-DMD for all but the second eigenvalue for the subcritical system with added synthetic noise, we do not interpret this result as evidence of superiority. It is not unreasonable to assume that an expert user of optimized-DMD could select more optimal parameters.  

\subsection{Monte Carlo transport criticality benchmark}\label{subsec:mcdc}
We present another neutronics example which is closer to practical engineering applications to test weak-DMD on data with non-synthetic noise. Kornreich and Parsons \cite{kornreich2005time} give analytic benchmark values for the first two time eigenvalues of a multi-region, monoenergetic, slab reactor problem. While the energy group structure is less sophisticated than the last problem, the inclusion of transport effects makes this problem analytically intractable and necessarily of much higher dimension.  


Because the time dependent solution is a function of angle and the $x$ coordinate, the problem is two dimensional. Neutron tallies in angle and space were calculated with the open source particle transport code MCDC \cite{morgan2024monte} and then flattened into a series of snapshots consisting of one dimensional vectors. For these calculations, the solution was calculated on a discretized domain of $182$ $x$ points and on a discrete angular grid formed by subdividing the cosine of the polar angle, $\mu$, in $196$ directions. Results were generated for $10^4$, $10^5$, and $10^6$ particle counts with $420$ unevenly spaced timesteps. Instead of dealing with neutron populations directly, tallies were used to calculate fluxes, which have units of energy per area per time. The angularly dependent flux is known as the angular flux, which is summed over angle to return the scalar flux. Snapshots over the range $t\in[\times10^{-6}, 10^{-1}]\:\mu\mathrm{s}$ were selected for DMD. 

\begin{figure}
    \centering
    \includegraphics[draft=false,width=0.75\linewidth]{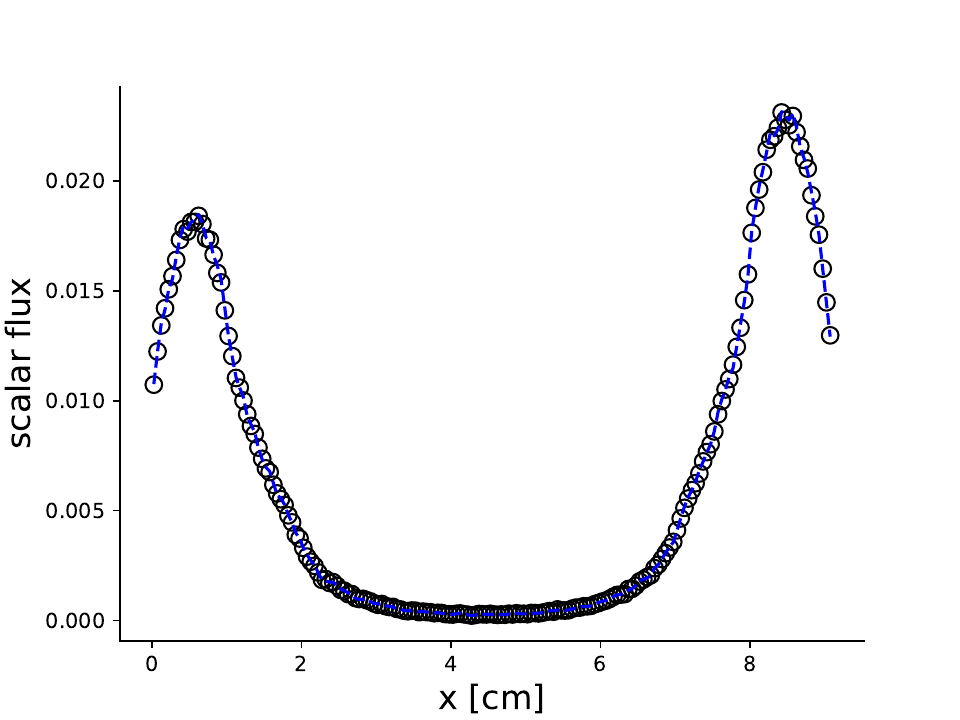}
    \caption{Scalar flux data (open circles) at $t=0.057\:\mu\mathrm{s}$ and reconstructed solution from weak-DMD with $12$ bases of order $p=3$ in the trial space for the Kornreich and Parsons benchmark problem.  }
    \label{fig:ex_phi_asymm-super}
\end{figure}

\begin{figure}[h!]
    \centering
    \begin{subfigure}[b]{0.49\textwidth}
        \includegraphics[draft=false,width=\textwidth]{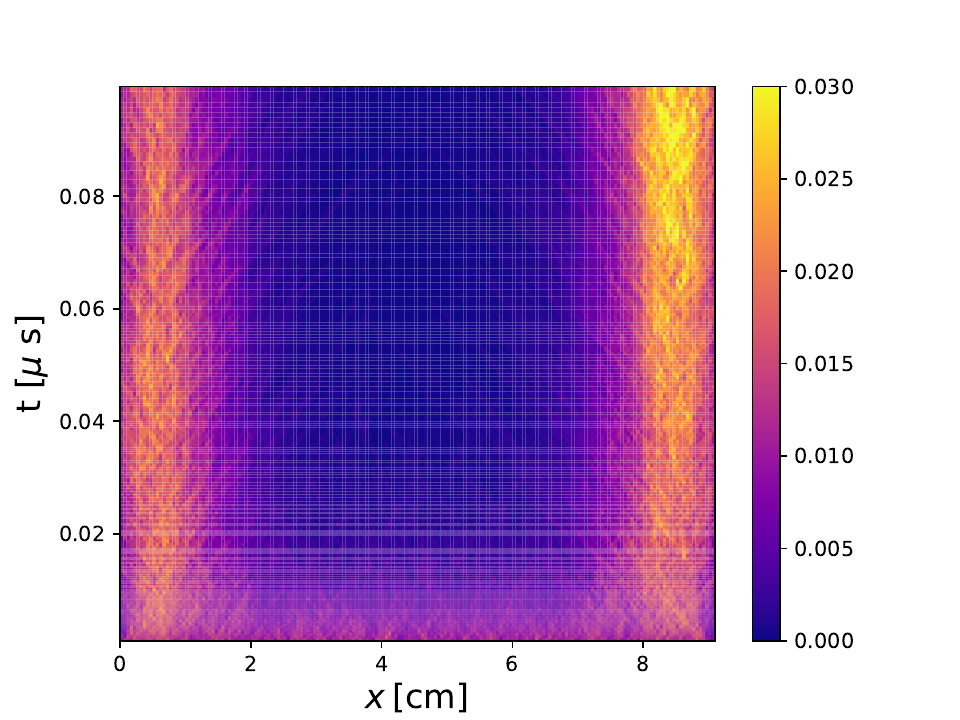}
        \caption{$10^4$ particles data}
    \end{subfigure}
    \hfill 
    \begin{subfigure}[b]{0.49\textwidth}
        \includegraphics[draft=false,width=\textwidth]{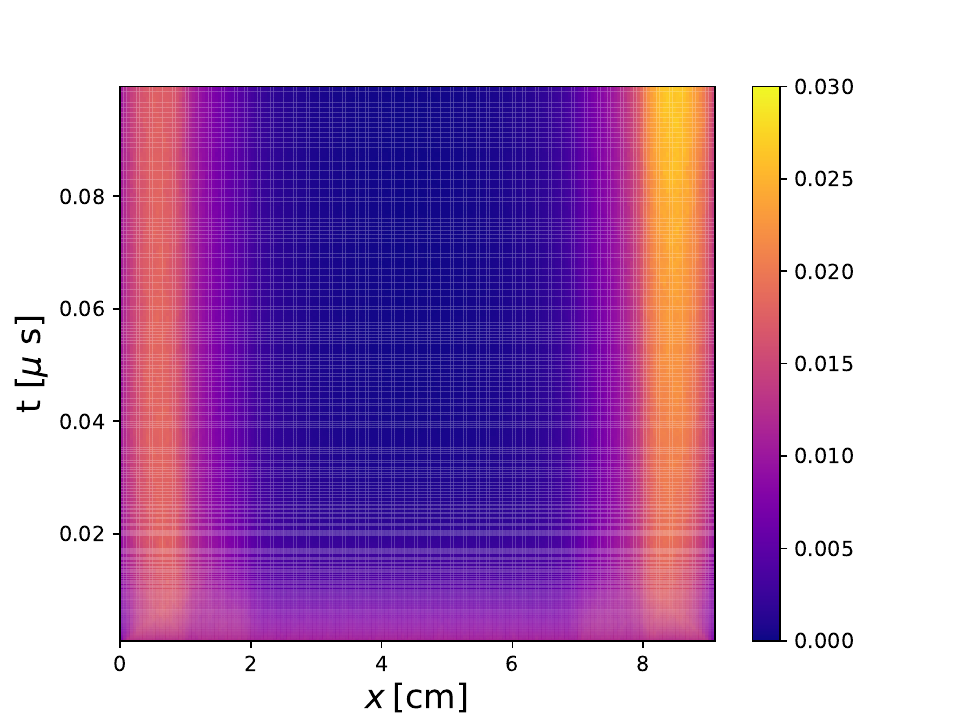}
        \caption{$10^6$ particles data}
    \end{subfigure}
    \begin{subfigure}[b]{0.49\textwidth}
        \includegraphics[draft=false,width=\textwidth]{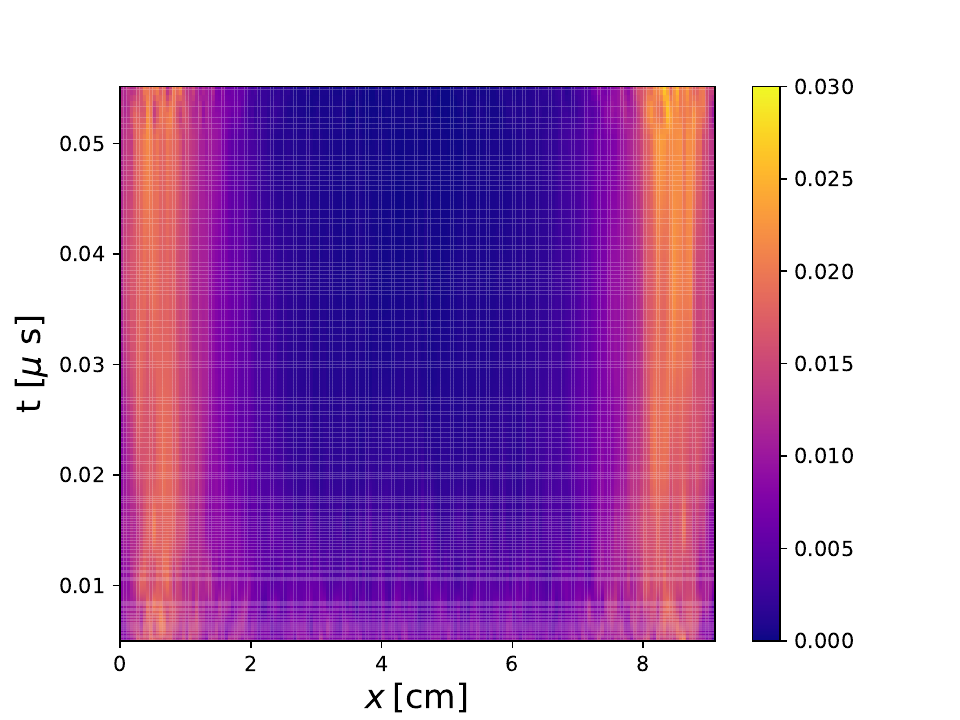}
        \caption{$10^4$ particles reconstruction}
    \end{subfigure}
    \begin{subfigure}[b]{0.49\textwidth}
        \includegraphics[draft=false,width=\textwidth]{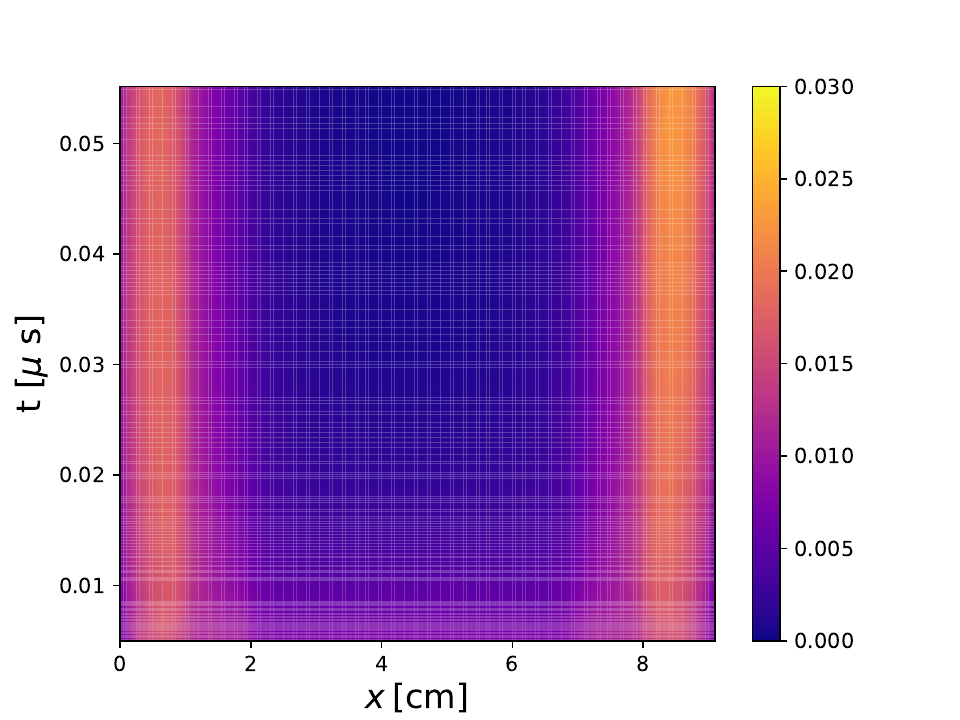}
        \caption{$10^6$ particles reconstruction}
    \end{subfigure}
    \caption{Color map of scalar flux data (top row) and reconstructions (lower row) for the Kornreich and Parsons criticality benchmark for $10^4$ (left column) and $10^6$ (right column) particles. Reconstructions were made by projecting onto a trial space of $12$ basis functions of order $p=3$.  }
    \label{fig:scalar_flux_2D_total}
\end{figure}

\begin{figure}[h!]
    \centering
    \begin{subfigure}[b]{0.49\textwidth}
        \includegraphics[draft=false,width=\textwidth]{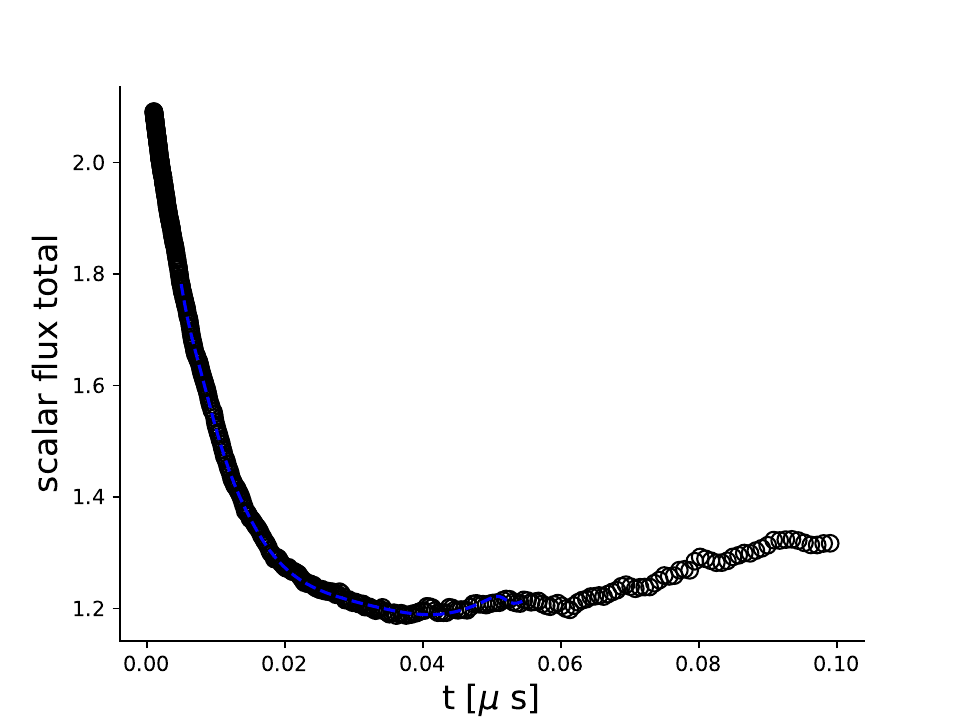}
        \caption{$10^4$ particles}
    \end{subfigure}
    \hfill 
    \begin{subfigure}[b]{0.49\textwidth}
        \includegraphics[draft=false,width=\textwidth]{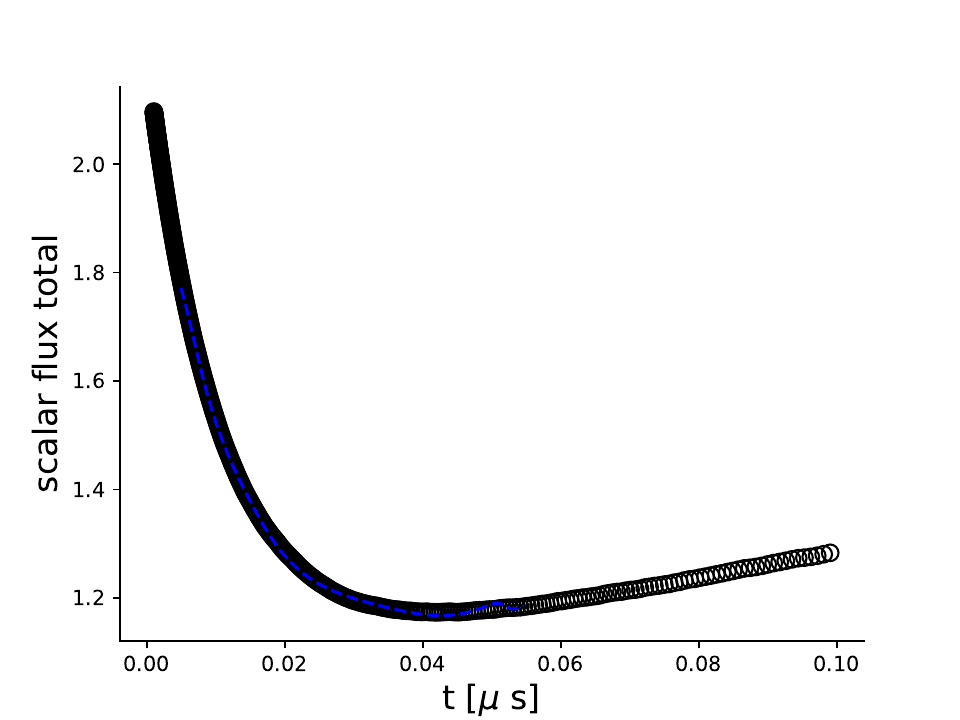}
        \caption{$10^6$ particles}
    \end{subfigure} 
    \label{fig:scalar_flux_sum}
    \caption{Sum total scalar flux in the slab reactor Kornreich and Parsons criticality benchmark. Blue dashed lines represent reconstructions and empty circles mark the data points. Reconstructions were made by projecting onto a trial space of $12$ basis functions of order $p=3$. }
\end{figure}

Unlike the results of Section \ref{subsec:12grp}, the noisiness of this data is not synthetic but arises naturally from statistical uncertainty inherent to the numerical method. The signal noise is proportional to the number of particles, which is evident in Figure \ref{fig:scalar_flux_2D_total}, which shows the time evolution of the scalar flux inside the domain with data plots and plots of the smoothed reconstructions. The scalar flux is calculated by summing the full solution over angle. Figure \ref{fig:ex_phi_asymm-super} gives a snapshot of the scalar flux data at $t\approx 0.05\:\mu\mathrm{s}$ as well as the reconstructed solution also summed over angle. The sum of the scalar flux over all spatial points is plotted in Figure \ref{fig:scalar_flux_sum}. Summing over angle has a noise canceling effect. From these figures, the solution characteristics become apparent. Two spikes in the neutron population occur in each fuel region and are decoupled by the moderator. The fuel region on the right becomes supercritical, while the region on the left does not. 

\begin{table}[]
\begin{tabular}{l|lllll}
Method        & Particles & Eigenvalue 1 & Error 1 & Eigenvalue 2      & Error 2 \\ \hline
weak-DMD      & 10000                  & (0.0251)  & 0.0125  & (-0.002+0.5$6\iu$)    & 0.5601  \\
weak-DMD      & 100000                 & (0.1483)  & 0.1107  & (-0.0437)      & 0.05    \\
weak-DMD      & 1000000                & (0.0154)  & 0.0222  & (-0.8283)      & 0.8346  \\
optimized-DMD & 10000                  & (-0.0153) & 0.0529  & (-1.4152-3.088$1\iu$) & 3.3997  \\
optimized-DMD & 100000                 & (-0.5375) & 0.5751  & (-0.5533)      & 0.5596  \\
optimized-DMD & 1000000                & (-0.4246) & 0.4622  & (-0.4259)      & 0.4322  \\
V-DMD         & 10000                  & (-0.057)  & 0.0946  & (-0.1815+1.529$7\iu$) & 1.5411  \\
V-DMD         & 100000                 & (0.0099)  & 0.0277  & (-0.1968+1.4361$\iu$) & 1.4504  \\
V-DMD         & 1000000                & (0.0177)  & 0.0199  & (-0.6234+0.349$1\iu$) & 0.7201 
\end{tabular}
\caption{Comparison of eigenvalues estimated by three DMD methods applied to MCDC data from the supercritical assembly of Kornreich. The benchmark eigenvalues are $\lambda_1= 0.03759991$, $\lambda_2 = -0.006298843$. For optimized-DMD, $8$ singular values were retained and results were averaged over $4$ trials. For the VDMD results, the energy of the retained singular values was $\mathcal{E}=0.4$. This parameter setting seemed to provide some resilience to noise. For weak-DMD, $12$ bases were employed in the trial reconstruction and $30$ in the test having polynomial order $p=3$ with $\mathcal{E}=0.91$.  }
\label{tab:MCDC-table}
\end{table}

For particle counts of $10^4$, $10^5$, and $10^6$, Table \ref{tab:MCDC-table} gives the first two leftmost eigenvalues returned by optimized-DMD, weak-DMD, and VDMD, as well as the the error. Error is calculated by
\begin{equation}
    \mathrm{Error} = \sqrt{\mathrm{Re}(\lambda-\tilde{\lambda})^2 + \mathrm{Im} (\lambda-\tilde{\lambda})^2},    
\end{equation}
where $\lambda$ is the benchmark eigenvalue and $\tilde{\lambda}$ is the DMD approximation. It is interesting to note that for a very low singular value energy tolerance ($\mathcal{E}=0.4$), VDMD is able to reasonably estimate the dominant mode for the $10^5$ and $10^6$ particle cases with lower errors than weak-DMD and optimized-DMD. The dominant modes from optimized-DMD from all particle counts are negative, which is incorrect for this supercritical assembly. The weak-DMD results are closer to the benchmark values than the optimized-DMD results for all but the second eigenvalue in the $10^6$ particle case. The data snapshots were sparse compared to the data of Section \ref{subsec:12grp}. Weak-DMD results seemed to benefit from a lower singular value energy tolerance and more peaked basis polynomials $p=3$.


\subsection{Flow past a cylinder}\label{subsec:cylinder}
Last, we apply weak-DMD and optimized-DMD to an \texttt{OpenFOAM} \cite{jasak2009openfoam} dataset used in \cite{han2022predicting} containing fluid velocities from simulated flow past a two-dimensional cylinder for Reynolds numbers ranging from $\mathrm{Re}=300$ to $\mathrm{Re}=1000$. It is common practice to test new DMD algorithms on turbulent flow data \cite{dawson2016characterizing,jovanovic2014sparsity}.  \cite{bagheri2013koopman} specifically recommended that optimized-DMD be applied to the flow past a cylinder problem at high Reynolds number. 

Snapshots starting from $t\in [0,200]\:[\mathrm{s}]$ were fed into the weak-DMD and optimized-DMD algorithms for the data across Reynolds number. Parameters were chosen for each in an attempt to minimize the forecast error for the $\mathrm{Re}=300$ case. 

\begin{figure}[h!]
    \centering
    \begin{subfigure}[b]{0.49\textwidth}
        \includegraphics[draft=false,width=\textwidth]{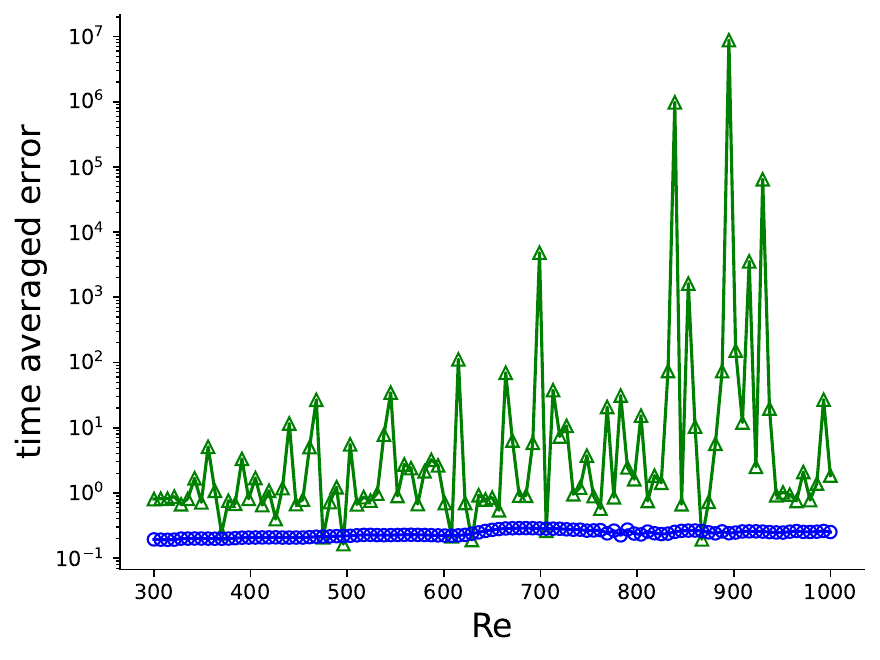}
        \caption{Averaged projection error}
    \end{subfigure}
    \hfill 
    \begin{subfigure}[b]{0.49\textwidth}
        \includegraphics[draft=false,width=\textwidth]{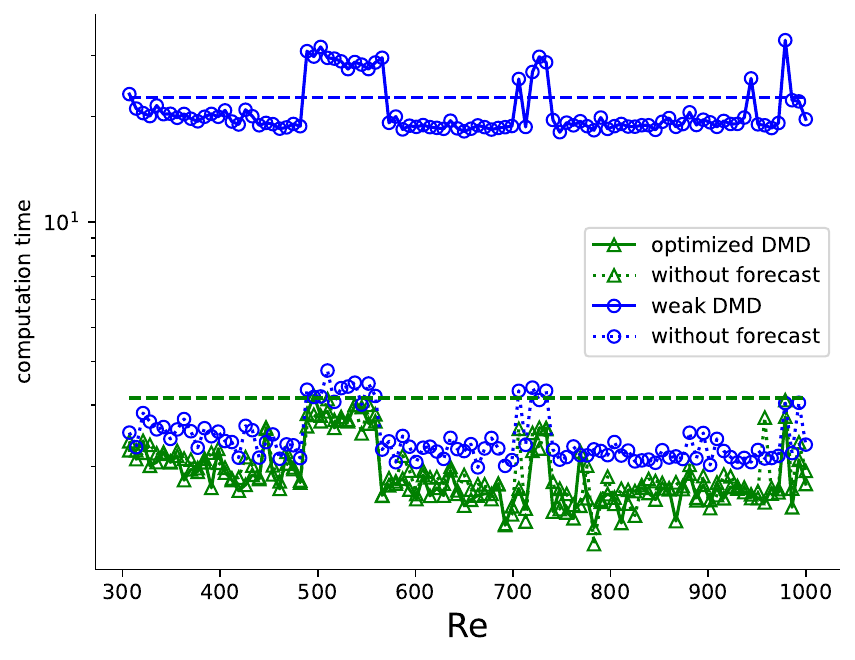}
        \caption{ Computational time}

    \end{subfigure}
    \caption{Log-log scaled averaged forecast error (Eq.~\eqref{eq:err_growth_cylinder}) for the flow past a cylinder dataset (a) and computational time (b) for each data-point in (a). In (a), the green line marked with triangles represents optimized-DMD results while the blue line with circles are weak-DMD results. In (b), The blue and green dashed lines respectively represent the instantiation times for weak-DMD and optimized-DMD. The dotted lines lines with symbols show the computational time with the time for the forecast step subtracted out. All experiments were run on a MacBook Pro equipped with a 2.8 GHz quad-core Intel Core i7 processor and 16 GB of RAM.}
    \label{fig:proj_err_time_cylinder}
\end{figure}

As analytic eigenvalues for the system are not available, error in the forecast was used to quantify performance. The time-dependent error for the forecast solution, $\widetilde{U}_x$, and the test data $U_x$, defined
\begin{equation}\label{eq:err_growth_cylinder}
    \mathrm{Error}(t) = \frac{||U_x(t)-\widetilde{U}_x(t)||_2}{||U_x(t)||},
\end{equation}
is averaged over $t\in[201, 250]\:[\mathrm{s}]$. This is a measure of the growth of the forecast error over time. The results of this test are plotted as a function of Reynolds number in Figure \ref{fig:proj_err_time_cylinder}, which also includes the computational times for each method.

\begin{figure}[htbp]
    \centering
    \begin{subfigure}[b]{0.48\textwidth}
        \centering
        \includegraphics[draft=false,width=\textwidth]{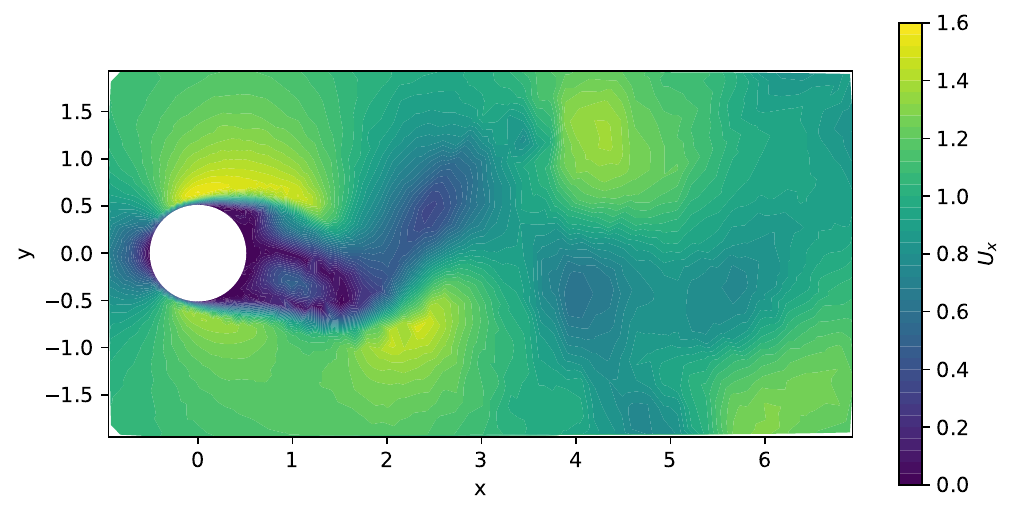}
        \caption{Velocity $x$ component $[250\:\mathrm{s}]$ into the simulation.}
    \end{subfigure}
    \hfill
    \begin{subfigure}[b]{0.50\textwidth}
        \centering
        \begin{subfigure}{0.32\textwidth}
            \centering
            \includegraphics[draft=false,width=\textwidth]{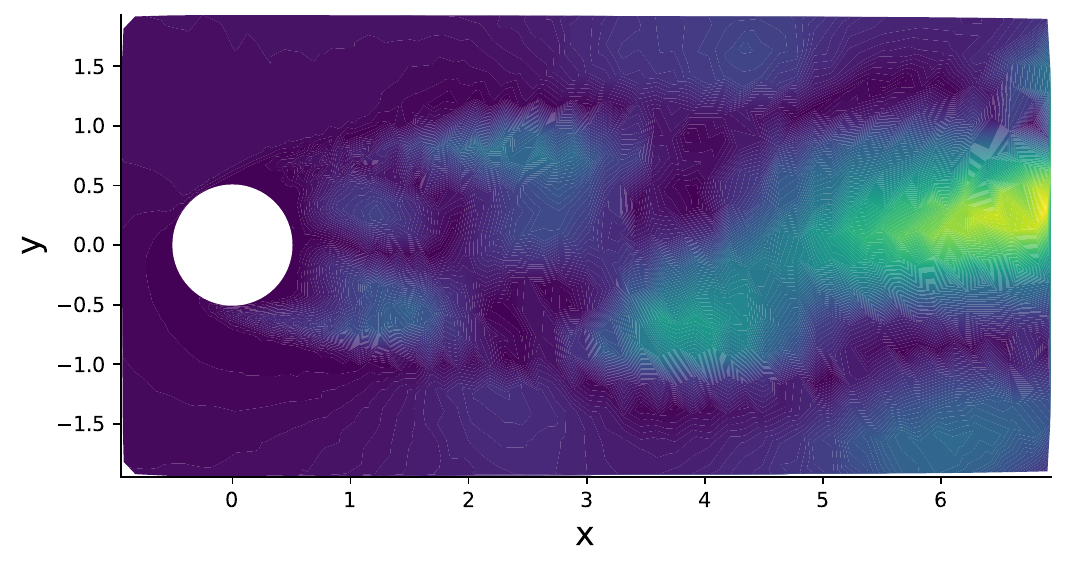}
            \caption{$k=0$}
        \end{subfigure}
        \hfill
        \begin{subfigure}{0.32\textwidth}
            \centering
            \includegraphics[draft=false,width=\textwidth]{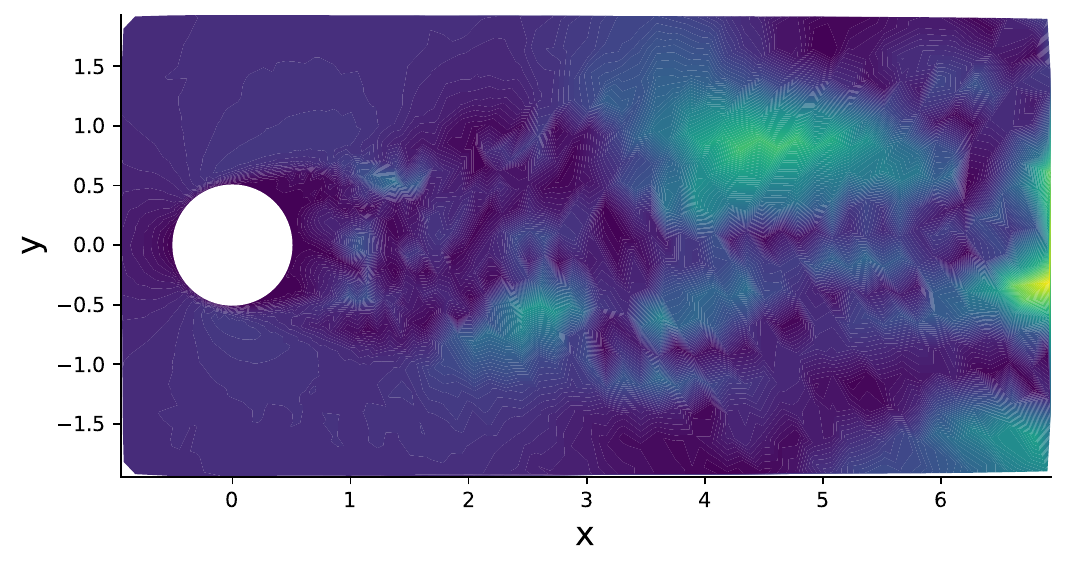}
            \caption{$k=2$}
        \end{subfigure}
        \hfill
        \begin{subfigure}{0.32\textwidth}
            \centering
            \includegraphics[draft=false,width=\textwidth]{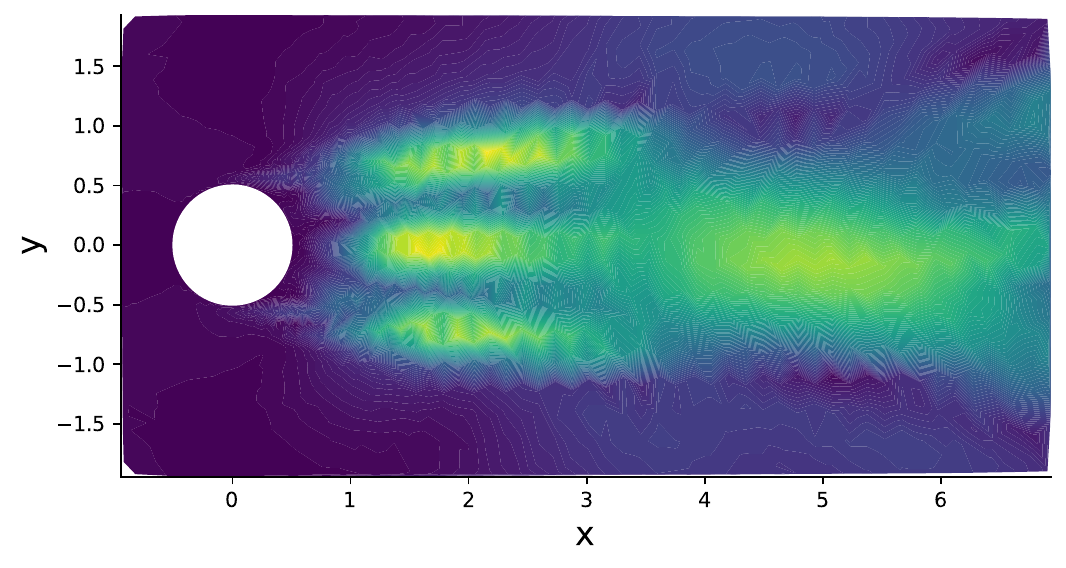}
            \caption{$k=4$}
        \end{subfigure}

        \vspace{0.6em}

        \begin{subfigure}{0.32\textwidth}
            \centering
            \includegraphics[draft=false,width=\textwidth]{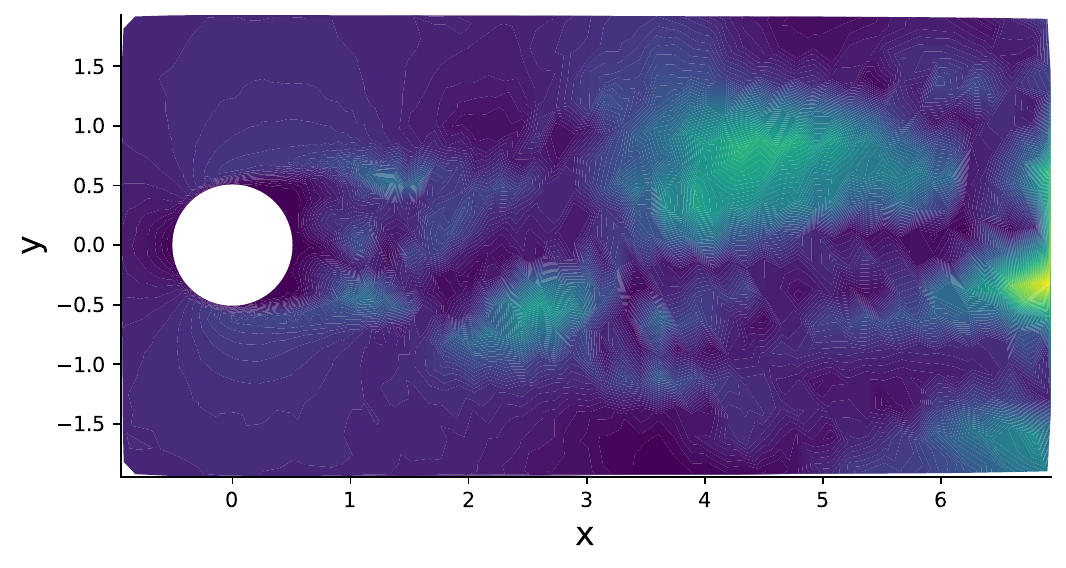}
            \caption{$k=1$}
        \end{subfigure}
        \hfill
        \begin{subfigure}{0.32\textwidth}
            \centering
            \includegraphics[draft=false,width=\textwidth]{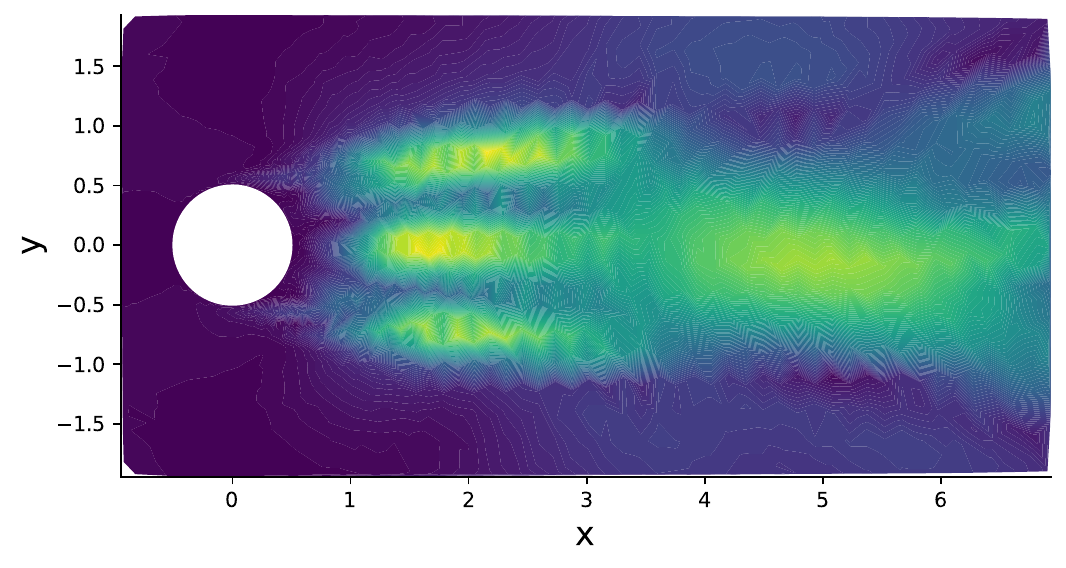}
            \caption{$k=3$}
        \end{subfigure}
        \hfill
        \begin{subfigure}{0.32\textwidth}
            \centering
            \includegraphics[draft=false,width=\textwidth]{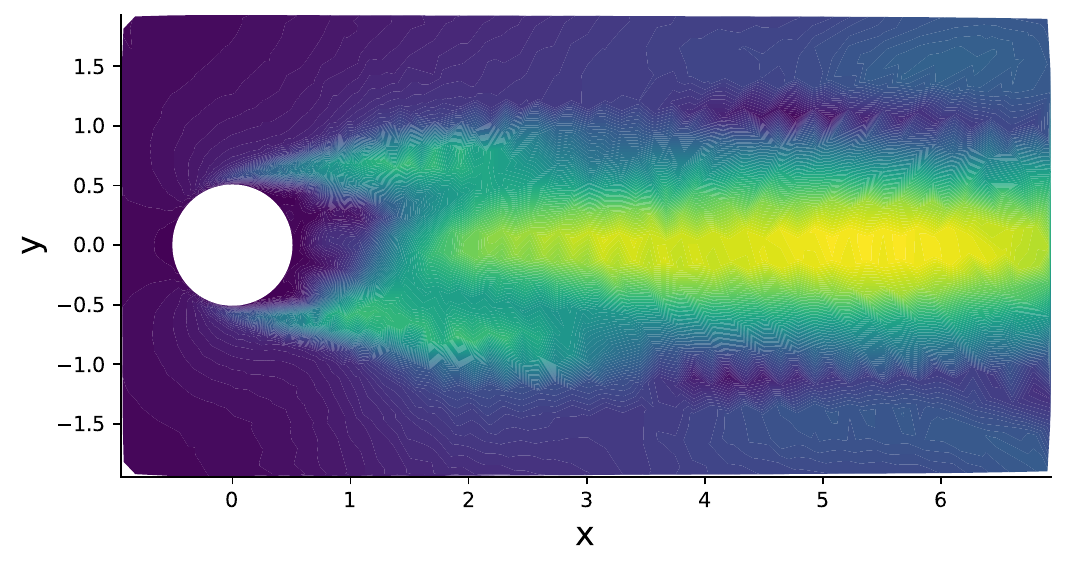}
            \caption{$k=5$}
        \end{subfigure}
    \end{subfigure}

    \caption{Data (a) and first six spatial modes normalized by the $L_2$ norm of the dominant mode ($k=0$) (b-g) estimated with weak-DMD for $\mathrm{Re} = 300$. The trial space included $17$ basis functions and the test, $48$ with $p=2$. The energy of the retained singular values was $\mathcal{E} = 0.9997$.  }
    \label{fig:modes+sol_Re300-weak}
\end{figure}

\begin{figure}[htbp]
    \centering
    \begin{subfigure}[b]{0.48\textwidth}
        \centering
        \includegraphics[draft=false,width=\textwidth]{new_figures/cylinder/solution_Re=300.0.pdf}
        \caption{Velocity $x$ component $[250\:\mathrm{s}]$ into the simulation.}
    \end{subfigure}
    \hfill
    \begin{subfigure}[b]{0.50\textwidth}
        \centering
        \begin{subfigure}{0.32\textwidth}
            \centering
            \includegraphics[draft=false,width=\textwidth]{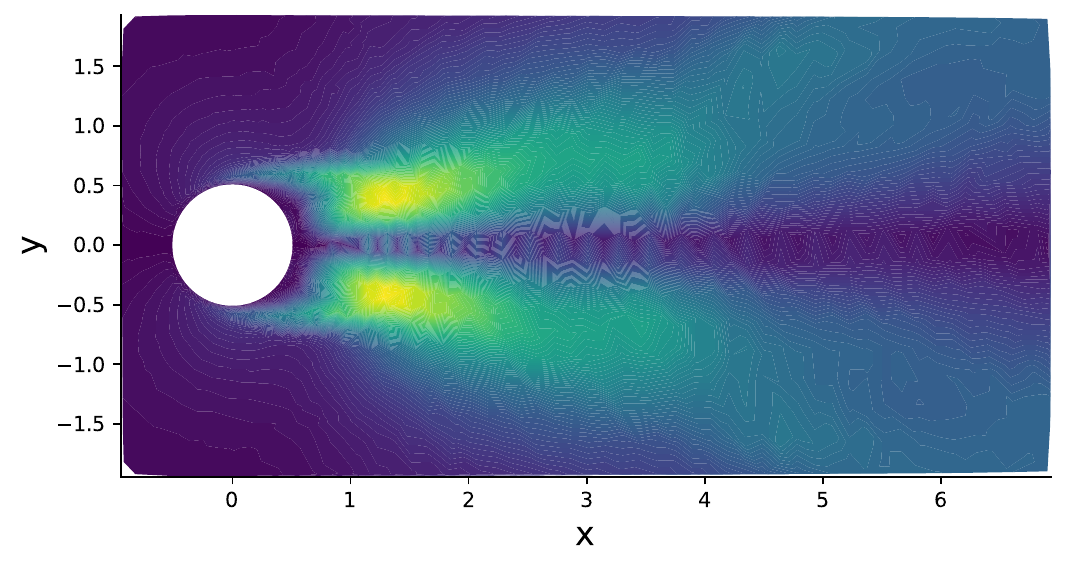}
            \caption{$k=0$}
        \end{subfigure}
        \hfill
        \begin{subfigure}{0.32\textwidth}
            \centering
            \includegraphics[draft=false,width=\textwidth]{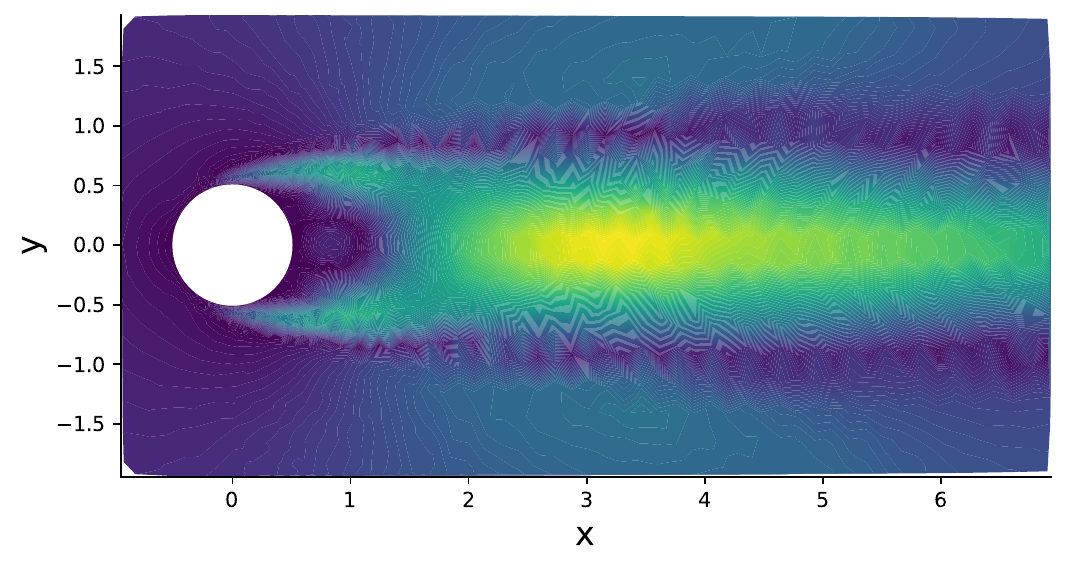}
            \caption{$k=2$}
        \end{subfigure}
        \hfill
        \begin{subfigure}{0.32\textwidth}
            \centering
            \includegraphics[draft=false,width=\textwidth]{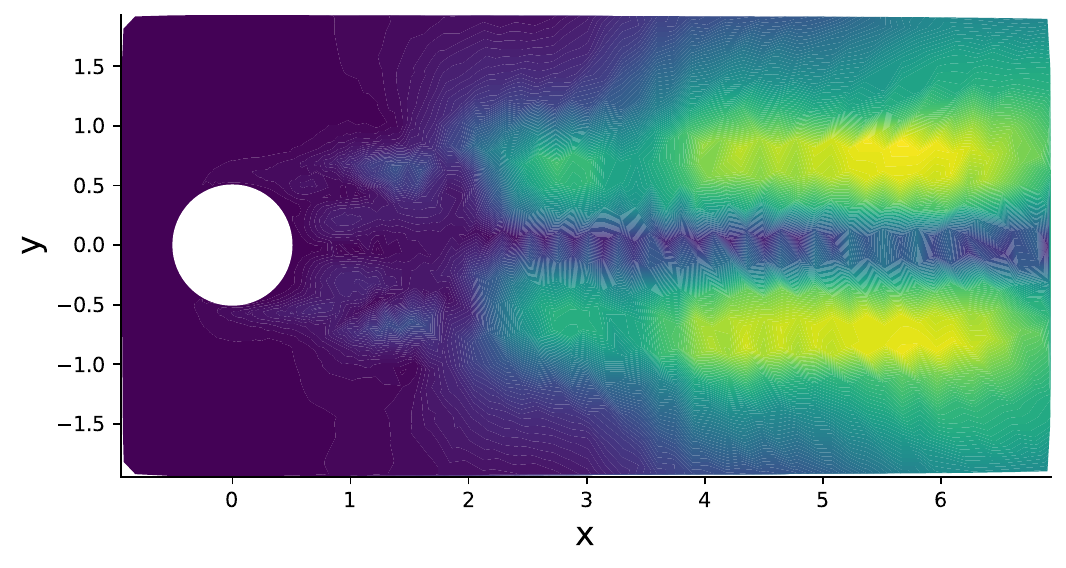}
            \caption{$k=4$}
        \end{subfigure}

        \vspace{0.6em}

        \begin{subfigure}{0.32\textwidth}
            \centering
            \includegraphics[draft=false,width=\textwidth]{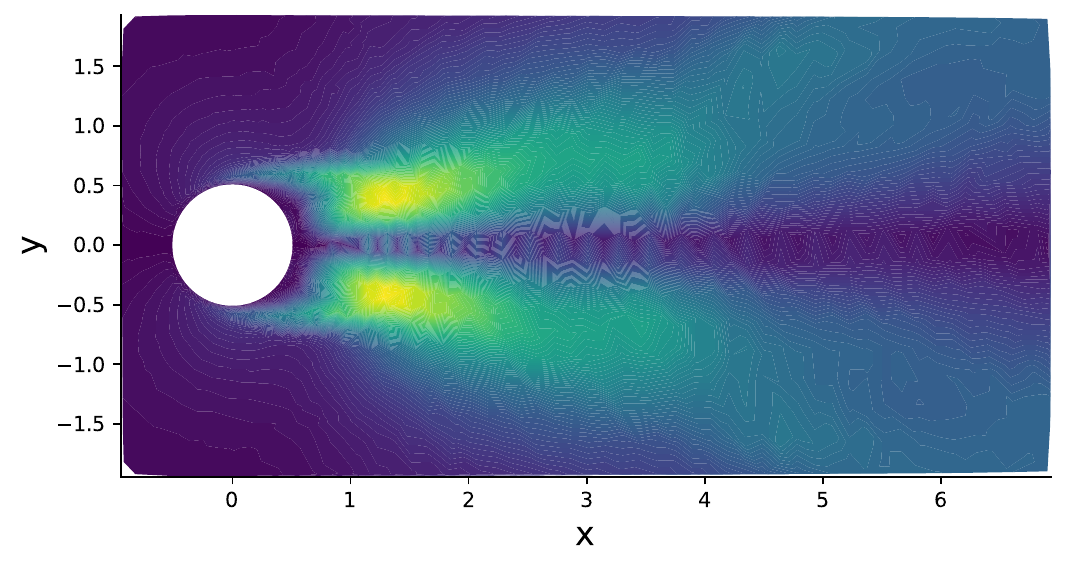}
            \caption{$k=1$}
        \end{subfigure}
        \hfill
        \begin{subfigure}{0.32\textwidth}
            \centering
            \includegraphics[draft=false,width=\textwidth]{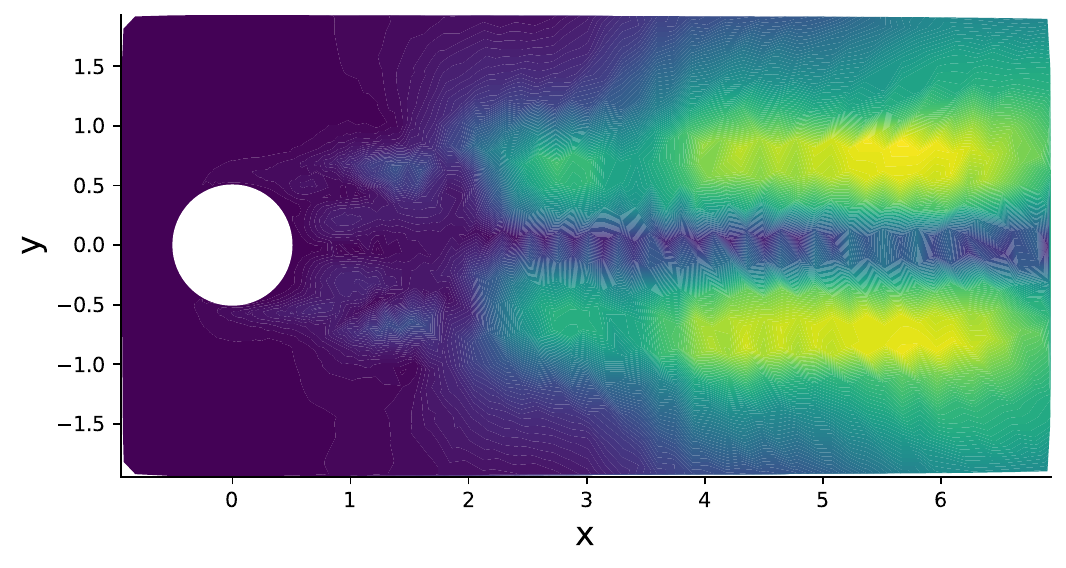}
            \caption{$k=3$}
        \end{subfigure}
        \hfill
        \begin{subfigure}{0.32\textwidth}
            \centering
            \includegraphics[draft=false,width=\textwidth]{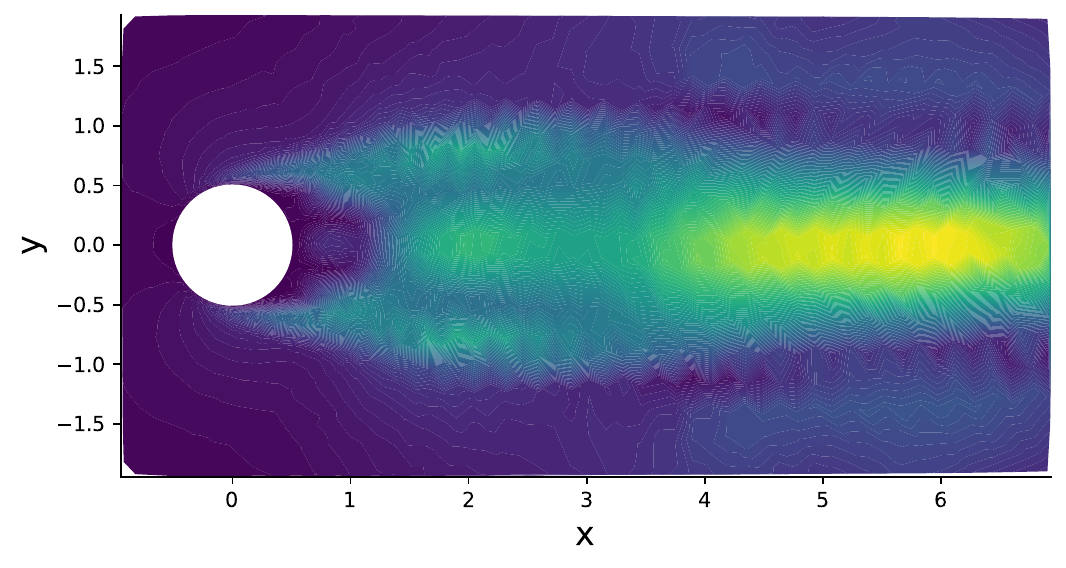}
            \caption{$k=5$}
        \end{subfigure}
    \end{subfigure}

    \caption{Data (a) and first six spatial modes normalized by the $L_2$ norm of the dominant mode ($k=0$) (b-g) estimated with optimized for $\mathrm{Re} = 300$. $8$ singular values were retained and results were averaged across $5$ trials.}
    \label{fig:modes+sol_Re300-opt}
\end{figure}

\begin{figure}[htbp]
    \centering
    \begin{subfigure}[b]{0.48\textwidth}
        \centering
        \includegraphics[draft=false,width=\textwidth]{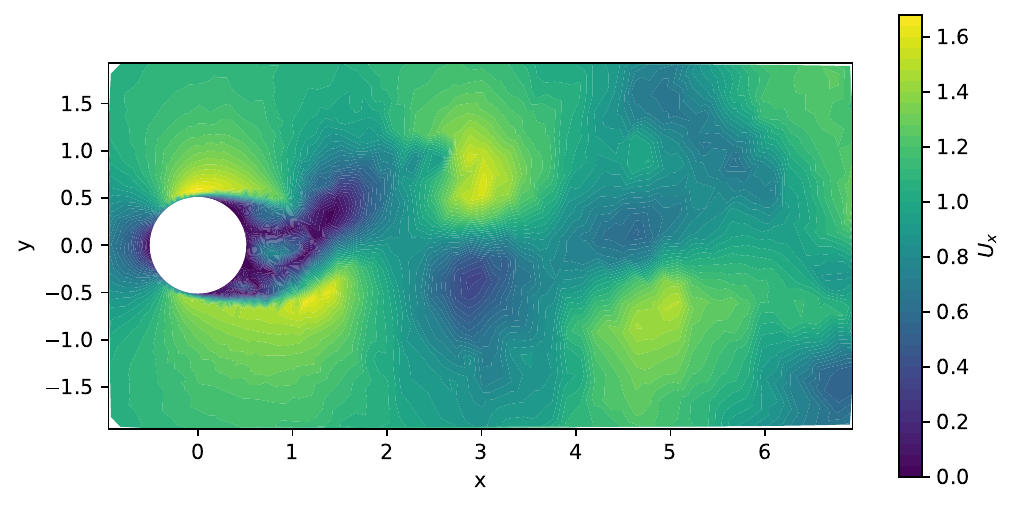}
        \caption{Velocity $x$ component $[250\:\mathrm{s}]$ into the simulation.}
    \end{subfigure}
    \hfill
    \begin{subfigure}[b]{0.50\textwidth}
        \centering
        \begin{subfigure}{0.32\textwidth}
            \centering
            \includegraphics[draft=false,width=\textwidth]{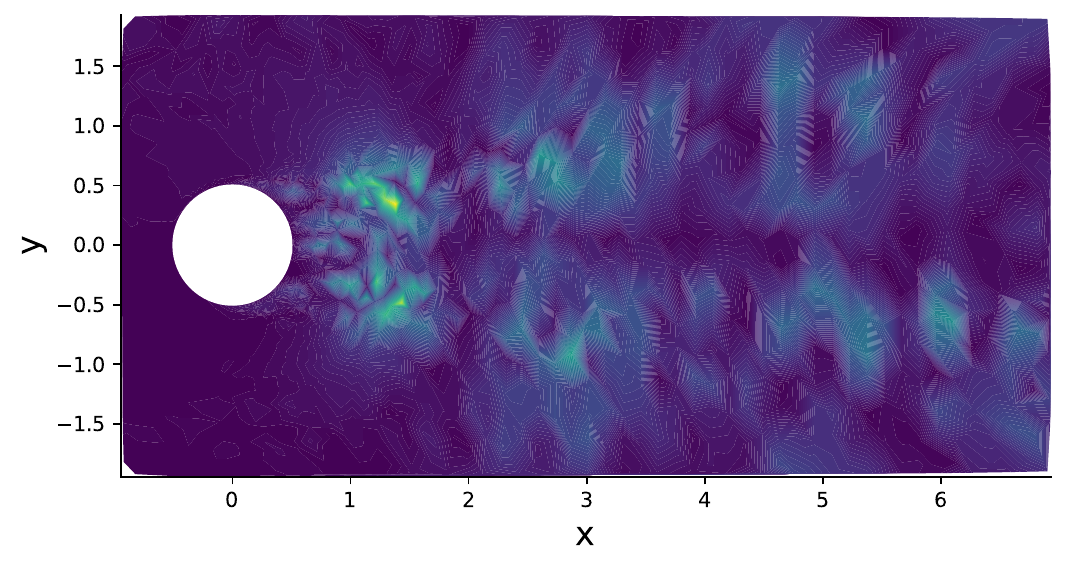}
            \caption{$k=0$}
        \end{subfigure}
        \hfill
        \begin{subfigure}{0.32\textwidth}
            \centering
            \includegraphics[draft=false,width=\textwidth]{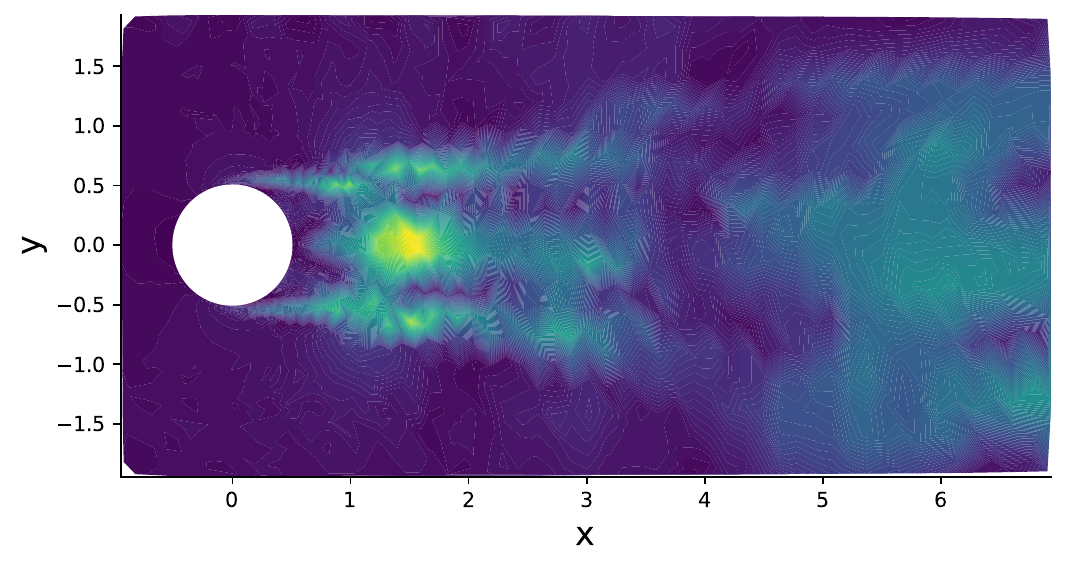}
            \caption{$k=2$}
        \end{subfigure}
        \hfill
        \begin{subfigure}{0.32\textwidth}
            \centering
            \includegraphics[draft=false,width=\textwidth]{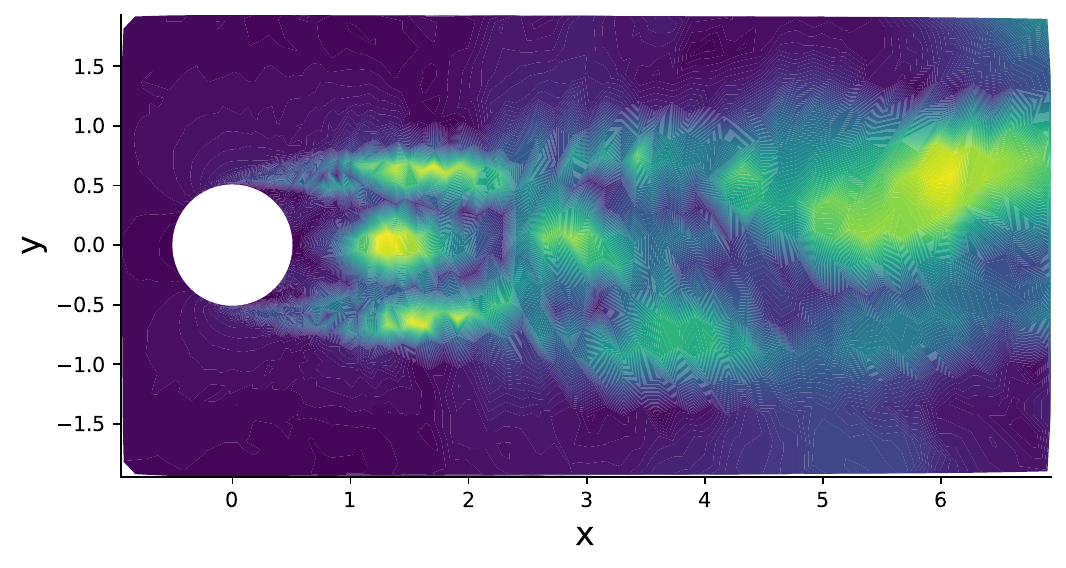}
            \caption{$k=4$}
        \end{subfigure}

        \vspace{0.6em}

        \begin{subfigure}{0.32\textwidth}
            \centering
            \includegraphics[draft=false,width=\textwidth]{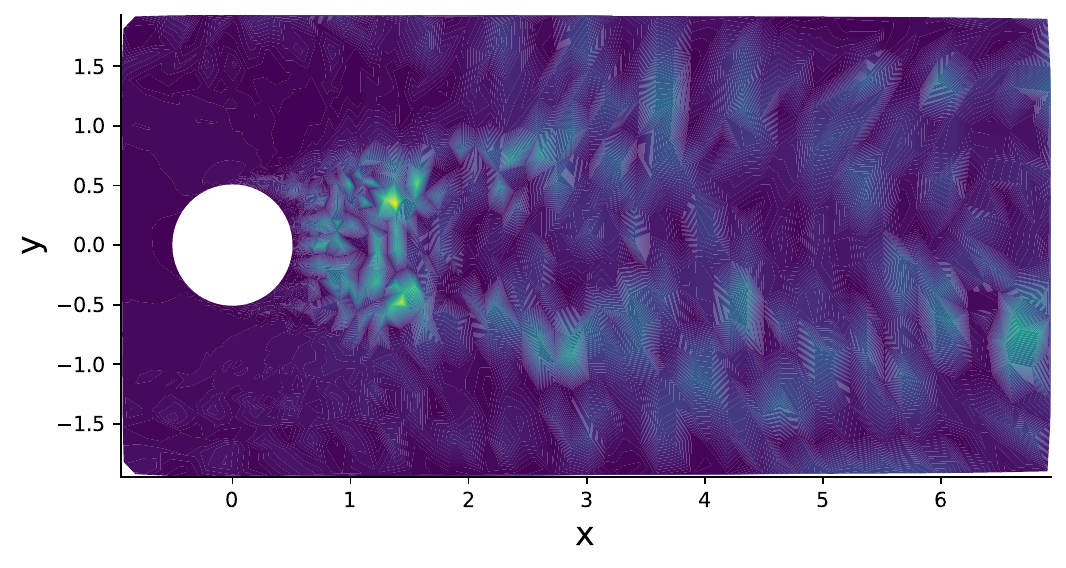}
            \caption{$k=1$}
        \end{subfigure}
        \hfill
        \begin{subfigure}{0.32\textwidth}
            \centering
            \includegraphics[draft=false,width=\textwidth]{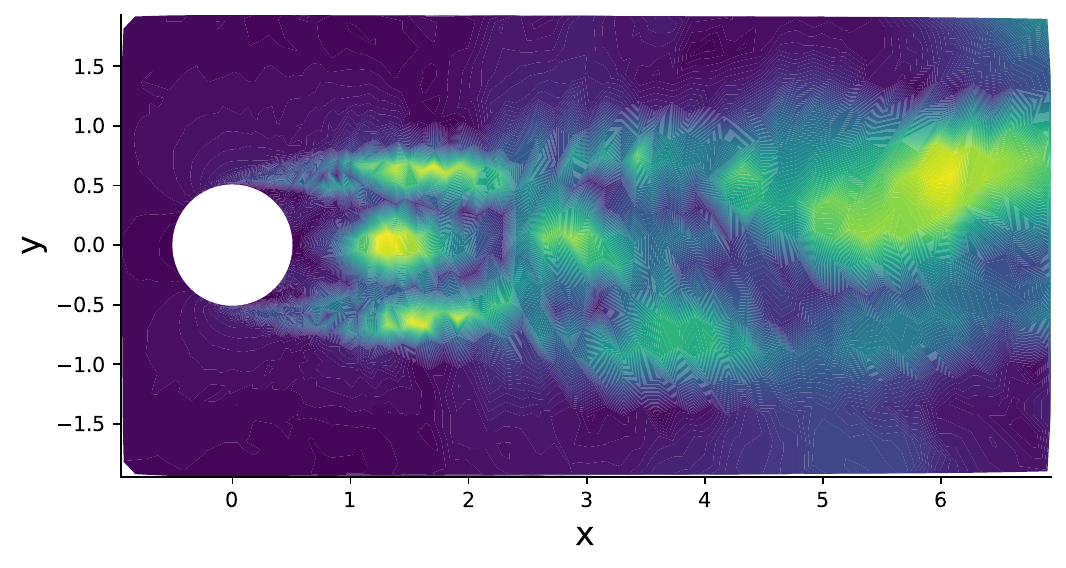}
            \caption{$k=3$}
        \end{subfigure}
        \hfill
        \begin{subfigure}{0.32\textwidth}
            \centering
            \includegraphics[draft=false,width=\textwidth]{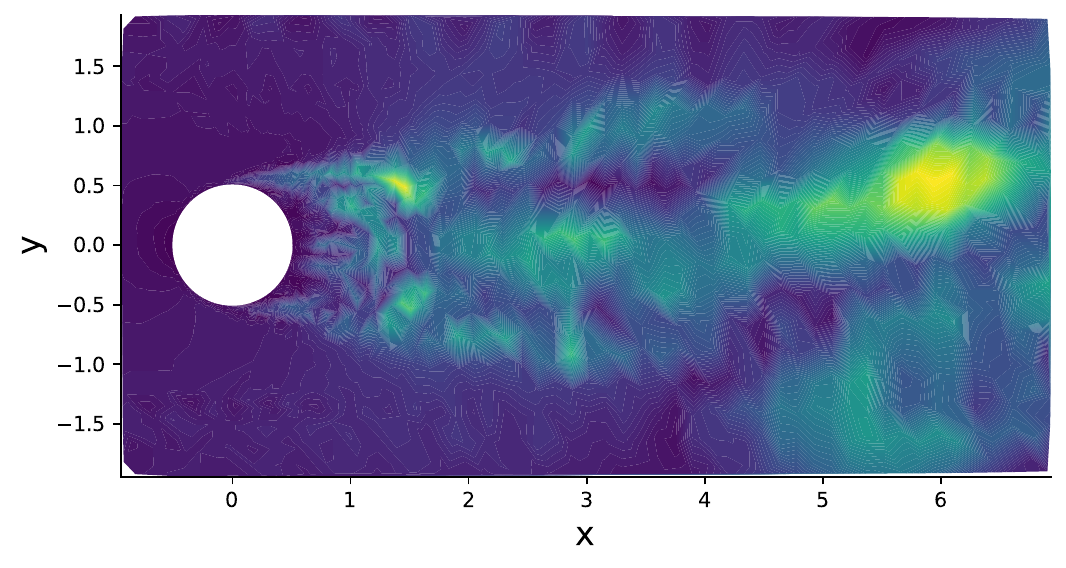}
            \caption{$k=5$}
        \end{subfigure}
    \end{subfigure}

    \caption{Data (a) and first six spatial modes normalized by the $L_2$ norm of the dominant mode ($k=0$) (b-g) estimated with weak-DMD for $\mathrm{Re} = 1000$. The trial space included $17$ basis functions and the test, $48$ with $p=2$. The energy of the retained singular values was $\mathcal{E} = 0.9997$.  }
    \label{fig:modes+sol_Re1000-weak}
\end{figure}

\begin{figure}[htbp]
    \centering
    \begin{subfigure}[b]{0.48\textwidth}
        \centering
        \includegraphics[draft=false,width=\textwidth]{new_figures/cylinder/solution_Re=1000.0.pdf}
        \caption{Velocity $x$ component $[250\:\mathrm{s}]$ into the simulation.}
    \end{subfigure}
    \hfill
    \begin{subfigure}[b]{0.50\textwidth}
        \centering
        \begin{subfigure}{0.32\textwidth}
            \centering
            \includegraphics[draft=false,width=\textwidth]{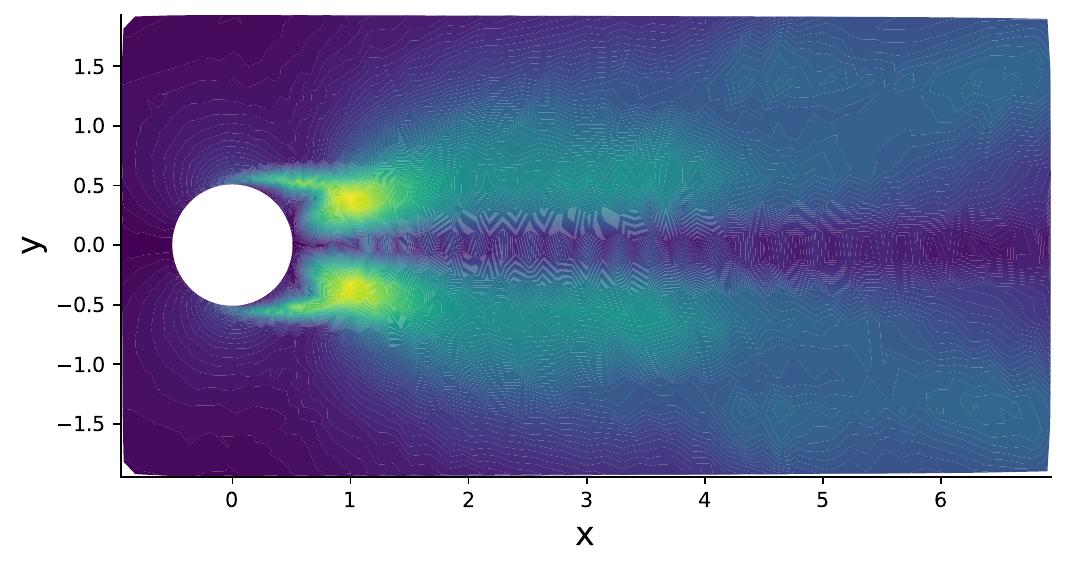}
            \caption{$k=0$}
        \end{subfigure}
        \hfill
        \begin{subfigure}{0.32\textwidth}
            \centering
            \includegraphics[draft=false,width=\textwidth]{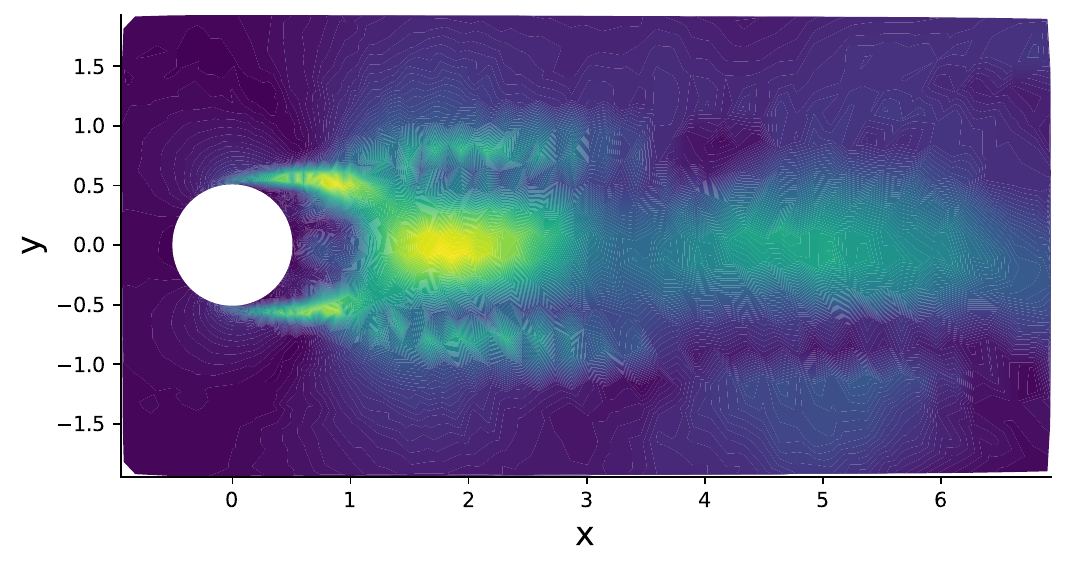}
            \caption{$k=2$}
        \end{subfigure}
        \hfill
        \begin{subfigure}{0.32\textwidth}
            \centering
            \includegraphics[draft=false,width=\textwidth]{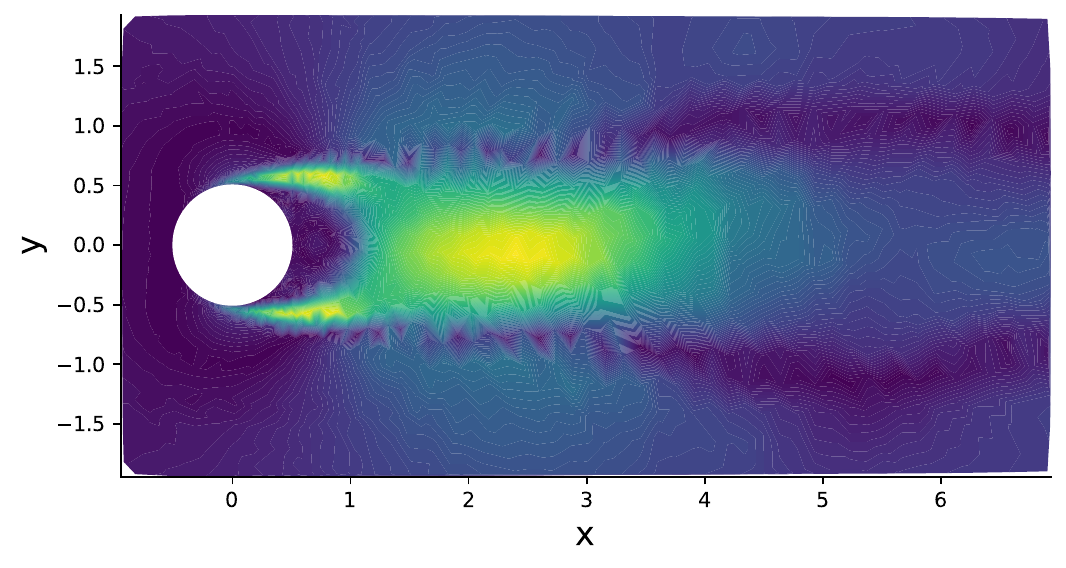}
            \caption{$k=4$}
        \end{subfigure}

        \vspace{0.6em}

        \begin{subfigure}{0.32\textwidth}
            \centering
            \includegraphics[draft=false,width=\textwidth]{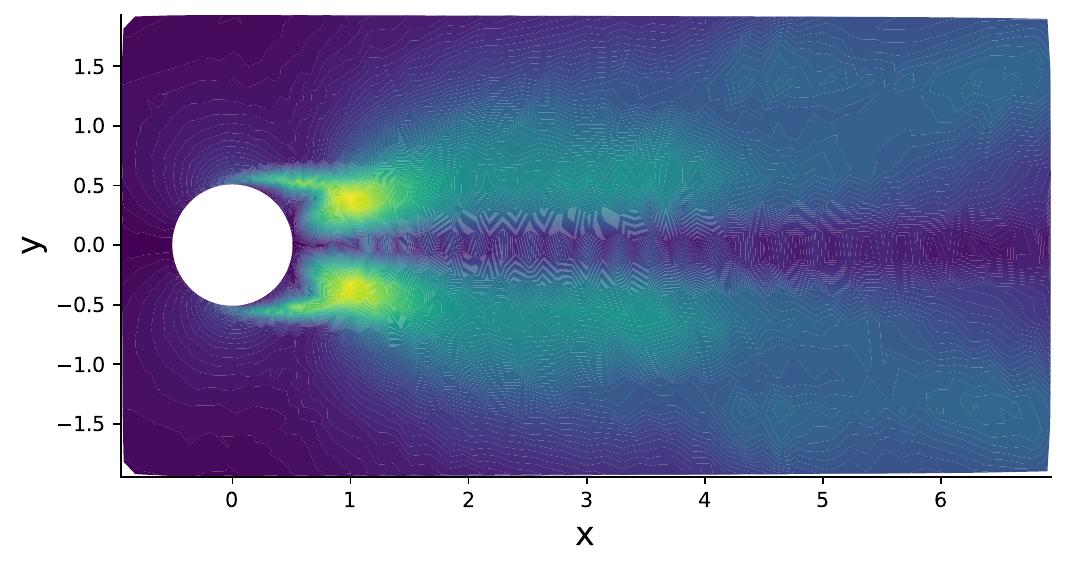}
            \caption{$k=1$}
        \end{subfigure}
        \hfill
        \begin{subfigure}{0.32\textwidth}
            \centering
            \includegraphics[draft=false,width=\textwidth]{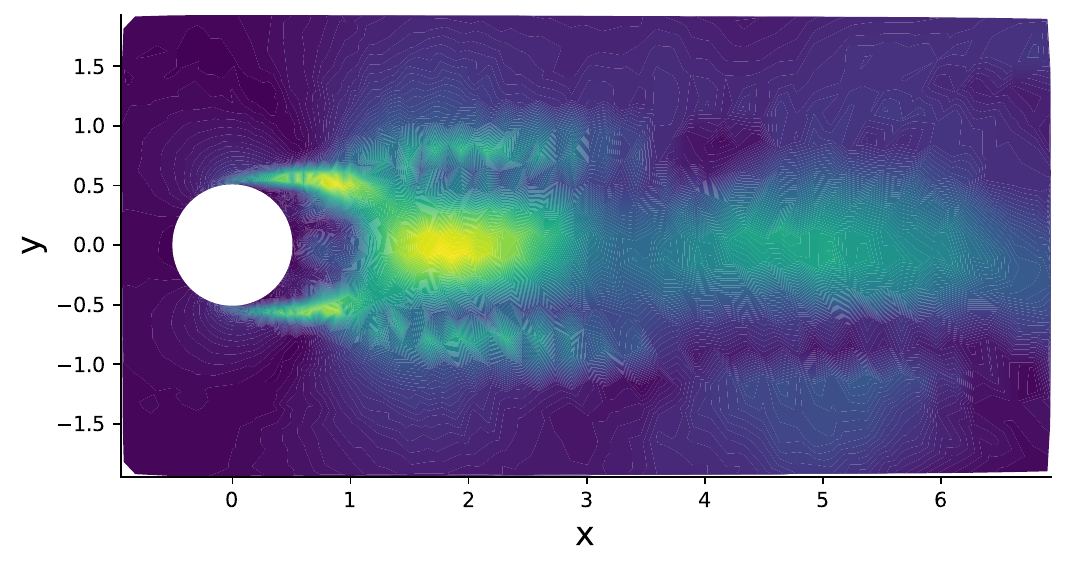}
            \caption{$k=3$}
        \end{subfigure}
        \hfill
        \begin{subfigure}{0.32\textwidth}
            \centering
            \includegraphics[draft=false,width=\textwidth]{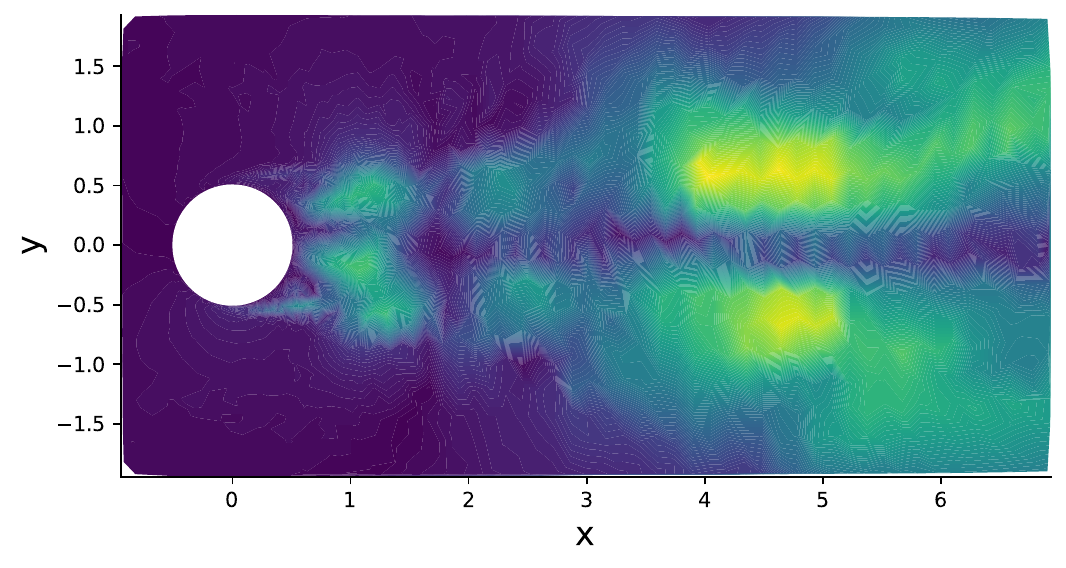}
            \caption{$k=5$}
        \end{subfigure}
    \end{subfigure}

    \caption{Data (a) and first six spatial modes normalized by the $L_2$ norm of the dominant mode ($k=0$) (b-g) estimated with optimized-DMD for $\mathrm{Re} = 1000$. $8$ singular values were retained and results were averaged across $5$ trials.  }
    \label{fig:modes+sol_Re1000-opt}
\end{figure}

Figures \ref{fig:modes+sol_Re300-opt} and \ref{fig:modes+sol_Re300-weak} show the data for the lowest Reynolds number as well as the first six modes calculated by optimized-DMD and weak-DMD respectively. Figures \ref{fig:modes+sol_Re1000-opt} and \ref{fig:modes+sol_Re1000-weak} do the same for the highest Reynold's number case. It is only possible to make qualitative comparisons from this data. Weak-DMD seems to be extracting more turbulent spatial solutions for the first few modes while optimized-DMD's dominant modes are structured. The quantitative results of Figure \ref{fig:proj_err_time_cylinder} show that the forecasts of optimized-DMD are less stable than those of weak-DMD with errors that are orders of magnitude greater than weak-DMD. Optimized-DMD is the clear winner in terms of computational cost, but removing the forecast step from both algorithms closes that gap significantly. This is a clear indication that the forecast step in weak-DMD is currently inefficient.


\section{Conclusions and future work}\label{sec:conc}
A Galerkin formulation of the DMD algorithm, weak-DMD, has been presented and compared with one of the more flexible and noise resilient methods extant, optimized-DMD. By casting the standard DMD system of ODEs into their weak form, restrictions on the time spacing of the data snapshots which inhibit many DMD methods are lifted. Furthermore, the noise reducing effects of integration onto a trial space serve to smooth noisy data. 

The method was demonstrated on three problems, a multi-energy group neutronics calculation in spherical geometry, a neutronics slab reactor benchmark, and the flow of incompressible fluid past a cylinder. The first two datasets had unevenly spaced timesteps. The $12$ group data was noised by sampling a Gaussian distribution, the slab reactor data contained noise inherent to the Monte Carlo method used to produce the data, and the cylinder flow data was not noised, but included deterministic randomness from turbulence.  

In terms of eigenvalue accuracy, weak-DMD was shown to perform comparably to optimized-DMD on the synthetic, $12$ group sphere problem, for which analytic eigenvalues are known. Weak-DMD correctly characterized the criticality slab reactor problem while optimized-DMD did not, and maintained a greater degree of forecast stability in the cylinder wake experiment. While optimized-DMD was generally faster and it is plausible that an expert user of optimized-DMD could improve the results, it can at least be argued that weak-DMD shows potential compared to the state of the art. 


In the current implementation of the weak-DMD method, there are many areas which invite improvement. The automation of the construction of trial and test basis spaces poses a significant challenge. A more user friendly implementation of weak-DMD would estimate the number, polynomial order, and support of each trial and test function from the characteristics of the data such as Fourier modes and estimated properties of the noise. The ever present question of the optimal number of retained singular values could also be approached in a manner similar to that of \cite{jovanovic2014sparsity}. 

Concerning analysis of the method, the relationship of the weak-DMD spectra to the spectra of the Koopman operator ought to be clarified. It will also be expedient to develop a theoretical understanding of how weak-DMD relates to nonlinear systems. Although the derivation assumed only measurement noise, the method was shown to perform well on a problem with inherent stochasticity in the state variables. Forgoing these analyses for now we present these initial results accompanied by assurance ``the proof of the pudding is in the eating.'' On the implementation side, certain rules of thumb emerged after analyzing the problems considered. It was generally better to have a larger test space than the trial space. \cite{messenger2021weak} also came to this conclusion. If the number of snapshots was small compared to the dimension of the system, retaining fewer singular values seemed to be expedient. These behaviors still merit further consideration. It remains to be seen whether other types of trial or test basis functions could improve method performance. The question of convergence, whether the algorithm will converge to a certain spectra in the limit of infinitely large test space, remains open.



\FloatBarrier
\appendix

\bibliography{ref.bib}
\end{document}